\documentclass[11pt]{article}
\usepackage{mathrsfs}
\usepackage{amsfonts}
\usepackage{amsmath}
\usepackage{amssymb}

\numberwithin{equation}{section}

\author{ \bf Yu Hou   \and  \bf Engui Fan\footnote{Corresponding
author and  e-mail address:
      faneg@fudan.edu.cn} \and  \bf Peng Zhao}
\date{   \small{ School of Mathematical Sciences, Institute of Mathematics \\
 and Key Laboratory of Mathematics for Nonlinear Science, \\ Fudan
University, Shanghai 200433, P.R. China}}
\title{\bf \Large{ An alternative approach to the quasi-Periodic solutions of the Hunter-Saxton hierarchy} }
\begin{document}
\maketitle

\begin{abstract}
 This paper is dedicated to provide
 the global solutions of algebro-geometric type for all the equations of a
 new commuting hierarchy containing the Hunter-Saxton (HS) equation.
 Our main tools include the zero curvature method to derive the HS
 hierarchy,
 the generalized Jacobian variety, the generalized Riemann theta
 function, the Weyl $m$-fucntions $m_\pm(x,t,z)$, and the pole motion obtained by
 solving an inverse problem for the
 Sturm-Liouville equation $L(\psi_1)=-\psi_1^{\prime
 \prime}=zy\psi_1$.
 Based on these tools and the theory of
 nonautonomous differential systems, topological dynamics and
 ergodic theory, the algebro-geometric solutions are obtained for the
 entire HS hierarchy.
\end{abstract}

\section{Introduction}
We study here the algebro-geometric solutions for all the equations
of a new commuting hierarchy containing the Hunter-Saxton (HS)
equation,
     \begin{equation}\label{1.1}
        u_{xxt}=-uu_{xxx}-2u_{x}u_{xx},
     \end{equation}
 where $u(x,t)$ is the function of spatial variable $x$ and time
 variable $t$.
 It arises in two different physical contexts in two nonequivalent
 variational forms \cite{1,2}. The first is shown to describe the propagation
 of weakly nonlinear orientation waves in a massive nematic liquid
 crystal director field \cite{1,2}. The second is shown to describe the high
 frequency limit of the Camassa-Holm (CH) equation \cite{10,11,32}
   \begin{equation}\label{1.2}
   u_t-u_{xxt}+3uu_x=2u_xu_{xx}+uu_{xxx},
   \end{equation}
 which was originally introduced in \cite{10,11} as model equation for shallow
 water waves, and obtained independently in \cite{31} with a bi-hamiltonian
 structure.

 The HS equation is a completely integrable system with a bi-hamiltonian
 structure and hence it possesses a Lax pair, an infinite family of commuting
 Hamiltonian flows, as well as an associated sequence of conservation laws,
 (Hunter and Zheng \cite{2}, Reyes \cite{20}). Traveling waves,
 asymptotic and piecewise smooth solutions of (\ref{1.1}) were described
 in Alber {\it et al} \cite{7,8,9}.
 The inverse scattering
 solutions have been obtained by Beals, Sattinger and Szmigielski \cite{19}.
 Recently, Lenells \cite{23}, \cite{24} and also Khesin and Misio{\l}ek \cite{22} pointed
 out that it describes the geodesic flow on the homogeneous space related to
 the Virasoro group. Bressan and Constantin \cite{25}, also Holden \cite{26}
 constructed a continuous semigroup
 of weak, dissipative solutions. Yin \cite{27} proved the local existence of strong
 solutions of the periodic HS equation and showed that all strong solutions-except
 space independent solutions-blow up in finite time. Gui, Liu and
 Zhu \cite{28} studied the wave-breaking phenomena and global existence.
 Furthermore, Morozov \cite{29}, Sakovich \cite{30} and Reyes \cite{20}, \cite{21}
 investigated (\ref{1.1}) from
 a geometric perspective. In \cite{16},  we constructed  algebro-geometric solutions of  the whole HS hierarchy by using
  polynomial recursive and spectral analysis method, due to Gesztesy [12].

 Quasi-periodic solutions (also called algebro-geometric solutions or
 finite gap solutions) of nonlinear equations were originally studied
 on the KdV equation based on the inverse spectral theory and
 algebro-geometric method developed by pioneers such as the authors
 in Refs.\cite{3}-\cite{9}, then in \cite{12,33}. Roughly speaking,
 the algebro-geometric approach consists of finding solutions of HS
 hierarchy, which are strictly connected to meromorphic functions
 defined on a Riemann surface, in the sense that the zeros and the
 poles of such functions completely determine the solution one is
 looking for. On the other hand, the algebro-geometric setting is
 directly linked to a linear differential operator of
 Sturm-Liouville type, in particular to its spectrum. In fact, it is
 well known that if $u(x,t)$ is a solution of HS equation, then
 there exists a positive density function $y(x,t)=u_{xx}/2$ such
 that the spectrum of the related Sturm-Liouville operator
     \begin{equation*}
      L\psi_1=\frac{d^2}{dx^2}\psi_1=-zy\psi_1
     \end{equation*}
 does not depend on $t$ \cite{18}. The study of isospectral  classes
 of $L$ leads to the determination of the constants of motion for
 the associated solution $u$ of HS hierarchy.

Motivated by Zampogni's work   \cite{14},   in this paper  we derive another hierarchy for HS, linked to the
 Sturm-Liouville operator $L$, by using a zero curvature method.
 This approach has  a significant difference from  our paper  \cite{16}:  the algebro-geometric setting of this paper
 is  derives from the spectral properties of $L$,
 and in particular from the properties of the Weyl $m$-functions. So
 we give more relevance to the spectral problem than to the
 associated zero curvature relation, i.e. the constants of motion
 derive directly from the isospectral classes of $L$ and not from
 the zero-curvature relation. At the same time, the study of the
 spectral properties of $L$ is carried out using methods of
 nonautonomous differential equations and the classical theory of
 algebraic curves.

 We pose the problem of finding all the densities $y(x)$ such that
 Hypotheses 2.4 hold. It turns out that this problem is equivalent
 to that of finding solutions $u(x,t)$ of the $r$-th order HS
 equation having as initial condition a solution of the stationary
 $n$-th order HS equation. Indeed, there is a surprising relation
 between the Weyl $m$-function of $L$ and the entries of the matrix
 satisfying the zero-curvature relation, and this fact enables us to find
 explicit algebro-geometric solutions for HS hierarchy, when the
 initial data lie in an isospectral class of $L$.

 The outline of this paper is as follows. In section 2, we review
 some recent developments in the theory of the inverse
 Sturm-Liouville problem. The material discussed here can be found
 in \cite{15,38}. We use methods of nonautonomous differential
 systems, topological dynamics and ergodic theory, to characterize
 all ``ergodic potentials $p,q,y:\mathbb{R} \rightarrow \mathbb{R}$"
 with $p,y>0$, which constitute a non autonomous differential system
 (\ref{2.1}). Such a characterization is carried out by considering
 the finite poles $P_1(x),\ldots,P_n(x)$ of the Weyl $m$-functions
 $m_\pm(z,x)$, when interpreted in a ``dynamical way".

 In section 3 and section 4,
 we derive the stationary HS hierarchy and the time-dependent HS
 hierarchy by using a zero curvature method, respectively.
 A solution of the $n$-th order stationary HS equation is the suitable
 initial condition for solving the $r$-th order time-dependent HS
 equation, for every fixed $n>r \in \mathbb{N}$.

 In section 5, we investigate the relation between the matrix $B_n$
 of the stationary formalism and the Weyl $m$-functions $m_\pm$ of
 the Sturm-Liouville system. This relation is the key to solving the
 HS
 equation. In particular, we can see that the entry $F_n$ of $B_n$
 completely determines the Weyl $m$-functions $m_\pm$, or better,
 their common meromorphic extension $M(P,x)$ to a Riemann surface
 $\mathcal{R}$. The zeros of $F_n$ (when viewed as a polynomial of
 degree $n$ in $z$) are exactly the poles $P_i(x)$ $(i=1,\ldots,n)$
 of the function $M(P,x)$. Based on this fact and the results of
 section 2, we give the expression for the solution of the $n$-th
 order stationary HS equation in (\ref{5.14}).

 In section 6, we devoted to the time-dependent formalism. We
 obtain the expression for the solution $u(x,t_r)$ of the $r$-th
 order HS equation in (\ref{6.44}). This solution is of class
 $C^\infty(\mathbb{R}^2)$, since the poles $P_i(x,t)$ have the same
 regularity properties. Moreover, we study the properties of the
 Weyl $m$-functions $m_\pm$. A Riccati type equation with respect to
 the time variable $t$ is obtained.

 In section 7, we investigate the $(x,t)$-motion of the poles
 $P_i(x,t)$. It turns out that such a motion can be made clear by
 considering its isomorphic image through a generalized Abel map on
 the generalized Jacobian variety of a Riemann surface $\mathcal{R}$
 of genus $n$. In particular, we will find that the motion on the
 generalized Jacobian variety can be described by using a basis
 $(d\omega_1,\ldots,d\omega_n)$ of holomorphic differentials on
 $\mathcal{R}$, plus a non-holomorphic differential $d\omega_0$, in
 such a way that the $x$-motion is confined to the non-holomorphic
 coordinate, and defines there a linear function $\omega_0(x)$,
 while the $t$-motion determines a triangular structure on
 the first $n-1$ of the holomorphic coordinates. The image of
 the $t$-motion through
 the generalized Abel map is determined by functions $\omega_i(x,t)$
 $(i=1,\ldots,n-1)$. The last holomorphic coordinate $\omega_n(x,t)$
 remains implicitly defined in such a way that the vector
 $(\omega_0(x),\omega_1(x,t),\ldots,\omega_n(x,t))$ is contained in
 a translate $\Upsilon_0$ of the zero locus of the generalized Riemann
 theta function $\Theta_0(z)$, for every $x,t\in\mathbb{R}$.
 In addition, we give a brief description of how
 we can build singular Riemann surfaces (called ``generalized
 Riemann surfaces"), and explain how the classical theory can be
 extended to these new objects.

 In section 8, we show that the elementary symmetric functions of
 $n$-tuples of distinct points in $\mathcal{R}$ can be expressed in
 terms of a generalized Riemann theta function. It turns out that
 the $j$-th symmetric function of points
 $P_1,\ldots,P_n\in\mathcal{R}$, depends on the partial derivative
 with respect to the coordinate $\omega_{n-j}$ of the logarithm of
 $w_0 (P_1(x,t_r),\ldots,P_n(x,t_r))$.
 As an application, we derive the Riemann
 theta function representation for the solution $u(x,t)$ of the HS
 hierarchy.

\section{Preliminaries}
 In this section, we give some important terminology and results
 about the Sturm-Liouville operator, in particular a
 dichotomy-theoretic approach to the inversion problem for such
 operator. A complete treatment of the results in this section
 can be found in \cite{15,38}.

 Let $A$ be a compact metric space and $\{\tau_x,x\in\mathbb{R}\}$ a
 family of homeomorphisms of $A$ such that
    \begin{enumerate}
 \item[(i)] $\tau_0(a)=a,$ for all $a\in A$;

 \item[(ii)] $\tau_{x+s}(a)=\tau_x \circ \tau_s(a),$ for every $a\in A$ and
 $x,s\in\mathbb{R}$;

 \item[(iii)] the map $\tau: A \times \mathbb{R} \rightarrow A:
 (a,x)\mapsto \tau_x(a)$ is continuous.
   \end{enumerate}
A family $\{\tau_x\}$ satisfying (i)-(iii) is called a {\it flow} on
$A$. We fix a $\{\tau_x\}$-ergodic measure $\mu$ on $A$, and assume
that $A$ is the topological support of $\mu$, i.e. $\mu(V)>0$ for
every open set $V\subset A.$ The triple $(A, \{\tau_x\},\mu)$ is
called a {\it stationary ergodic process}.

Let $p,q,y: A \rightarrow \mathbb{R}$ be continuous functions with
$p,y$ strictly positive; for every $a \in A$, we consider the map
(denoted by $a(x)$, with abuse of notation),
   \begin{equation*}
   x \mapsto
   \left(
     \begin{array}{cc}
       0 & 1/p(\tau_x(a)) \\
       q(\tau_x(a))-zy(\tau_x(a)) & 0 \\
     \end{array}
   \right).
   \end{equation*}
We study the family of differential equations
   \begin{equation}\label{2.1}
   \left(
     \begin{array}{c}
       \psi_1 \\
       \psi_2 \\
     \end{array}
   \right)^\prime
   =\left(
     \begin{array}{cc}
       0 & 1/p(\tau_x(a)) \\
       q(\tau_x(a))-zy(\tau_x(a)) & 0 \\
     \end{array}
   \right)
   \left(
     \begin{array}{c}
       \psi_1 \\
       \psi_2 \\
     \end{array}
   \right),
   \quad a\in A,~z\in\mathbb{C},
   \end{equation}
which is equivalent to $-(p\psi_1^\prime)^\prime +q\psi_1=zy\psi_1$,
if the map $a \mapsto \frac{d}{dx}p(\tau_x(a))|_{x=0}$ is well
defined and continuous.

Let $\Phi_a(x)$ be the fundamental matrix solution for (\ref{2.1}).

\newtheorem{defi2.1}{Definition}[section]
 \begin{defi2.1}
 {\rm
 The family of equations $(\ref{2.1})$ is said to have an {\it
 exponential dichotomy} over $A$, if there are positive constants
 $K$, $\gamma$ and a continuous function $a \mapsto P_a : A
 \rightarrow \mathcal{P}=$ \{linear projections $P: \mathbb{C}^2
 \rightarrow  \mathbb{C}^2$\}
 such that the following estimates hold:
  }
 \end{defi2.1}
 \begin{enumerate}
 \item[(i)] $ \| \Phi_a(x)P_a\Phi_a(s)^{-1} \| \leq K e^{-\gamma(x-s)},
      \quad x \geq s,$

 \item[(ii)] $\| \Phi_a(x)(I-P_a)\Phi_a(s)^{-1} \| \leq  K e^{\gamma(x-s)},
       \quad x\leq s$.
 \end{enumerate}
It follows from the fact that {\it tr}$(a(x))$=0 for all
$x\in\mathbb{R}$ that both the image Im$P_a$ and the kernel Ker$P_a$
are one-dimensional, i.e., can be viewed as complex lines in
$\mathbb{C}^2$. The following proposition shows an important
characterization of the spectrum of (\ref{2.1}) for $\mu$-a.e. $a\in
A$. A proof can be found in \cite{15}.

\newtheorem{prop2.2}[defi2.1]{Proposition}
 \begin{prop2.2}
  For $\mu$-a.e. $a\in A$, the spectrum $\Sigma_a$ of $(\ref{2.1})$
  equals a closed set $\Sigma \subset \mathbb{R}$ which does not
  depend on the choice of $a.$ Moreover,
   $$ \mathbb{C} \setminus \Sigma =\{ z\in \mathbb{C},~
   (\ref{2.1})~ \textrm{has an exponential dichotomy}\}.$$
 \end{prop2.2}

Let now $a\in A$ and $z \in \mathbb{C} \setminus \Sigma$. We define
$m_\pm(a,z)$ to be the unique complex numbers such that Im
$P_a=Span(1,m_+(a,z))^T$ and Ker $P_a=Span$ $(1,m_-(a,z))^T$. As
usual, we set $m_+=\infty$ (or $m_-=\infty$) if Im $P_a=Span(0,1)^T$
(or Ker $P_a=Span(0,1)^T$). Moreover, it is easily proved that $\Im
m_+ \Im z>0$ and $\Im m_- \Im z<0$ for every $z$ with $\Im z \neq
0$. For every fixed $a \in A$, the maps $z \mapsto m_\pm(a,z)$ are
analytic in $\mathbb{C} \setminus \mathbb{R}$, and are meromorphic
in $\mathbb{C} \setminus \Sigma$. They coincide with the classical
Weyl $m$-functions (the definition and the properties of the
classical Weyl $m$-functions, see \cite{35,36}). For every
$z\in\mathbb{C}$, let $\beta(z)$ be the upper Lyapunov exponent for
(\ref{2.1}).

\newtheorem{prop2.3}[defi2.1]{Proposition}
 \begin{prop2.3}
 Fix $a\in A$. Let $I \subset \mathbb{R}$ be an open interval and
 suppose that $\beta(z)=0$ for a.e. $z\in I$. Then $m_\pm(a,z)$
 extend holomorphically through $I$. If $h_\pm$ denote the
 extensions of $m_\pm$, then
  \begin{equation*}
  h_+(z)=
  \begin{cases}
   m_+(a,z), & \Im z>0 \\
   m_-(a,z), & \Im z<0
  \end{cases}
  \quad \textrm{and} \quad
  h_-(z)=
  \begin{cases}
   m_-(a,z), & \Im z>0 \\
   m_+(a,z), & \Im z<0.
  \end{cases}
  \end{equation*}
 \end{prop2.3}
 A proof of this proposition can be found in \cite{15}.

Throughout all the paper we make the following fundamental
assumptions.

\newtheorem{hyp2.4}[defi2.1]{Hypotheses}
 \begin{hyp2.4}
   \begin{enumerate}
  \item[(H1)] The spectrum $\Sigma$ of $(\ref{2.1})$ is a finite
 union of intervals, i.e. $ \Sigma=[z_2,z_3] \cup \ldots \cup [z_{2n},\infty);$
  \item[(H2)] The Lyapunov exponent $\beta(z)$ vanishes a.e. in the
  spectrum, i.e. $\beta(z)=0$ for a.e. $z\in \Sigma$.
 \end{enumerate}
 \end{hyp2.4}

Now we give the algebro-geometric structure of this paper. The
results stated following are proved in \cite{38}. We first look for
formula of algebro-geometric type for the functions $p,q$ and $y$.
Consider the Riemann surface $\mathcal{R}$ described by the
algebraic relation
   \begin{equation*}
   w^2=z\prod_{m=1}^{2n}(z-z_m).
   \end{equation*}
As usual, $\mathcal{R}$ is obtained by the union of two Riemann
spheres $\mathbb{C}^2 \cup \{\infty\}$ cut open along $\Sigma$ and
glued together in the usual way, with genus $n$ and exactly $2n+2$
ramification points, namely $0, z_1,\ldots,z_{2n}, \infty$. Let
$\pi:\mathcal{R} \rightarrow \mathbb{C} \cup \{ \infty\}$ be the
canonical projection and $k(P)$ be the meromorphic function on
$\mathcal{R}$ defined by
   \begin{equation*}
   k(z)=\sqrt{(z-z_1)\ldots(z-z_{2n})},
   \qquad \pi(P)=z.
   \end{equation*}
Then both $m_+(a,\cdot)$ and $m_-(a,\cdot)$ define a single
meromorphic function $M_a(\cdot)$ on $\mathcal{R}$, such that
$M_a=m_+$ and $M_a \circ \sigma=m_-$, where $\sigma$ is the map
interchanging the sheets. $M_a(P)$ satisfies the Riccati equation
   \begin{equation*}
    M_x+z^{1/2}\frac{1}{p}M^2=z^{-1/2}q-z^{1/2}y,
    \qquad \pi(P)=z.
   \end{equation*}
The function $M_a$ has exactly $n$ finite poles
$P_1(x),\ldots,P_n(x)$, one for each interval in the resolvent set
$\mathbb{R} \setminus \Sigma$. For the properties of the function
$M_a$ and a complete discussion about its poles and the
algebro-geometric setting, see \cite{38}.

If we put $\displaystyle T(z)=\sum_{i=1}^n (z-\pi(P_i(x))),$ then we
have
   \begin{equation}\label{2.2}
    \begin{split}
    M_a(P)=m_+(a,P)=\frac{Q(z)+\sqrt{py}k(P)}{T(z)},
    \quad \pi(P)=z \\
    M_a \circ \sigma(P)=m_-(a,P)=\frac{Q(z)-\sqrt{py}k(P)}{T(z)},
    \quad \pi(P)=z,
    \end{split}
   \end{equation}
where $ Q(z)=\frac{i(py)^\prime}{4y}z^{n-1/2}
+q_{n-1}z^{n-3/2}+\ldots+q_0(x)$, and $Q(\pi(P_i))=\sqrt{py}k(P_i)$,
for every $i=1,\ldots,n$.

The function $M_a$ can be viewed as a function of the variable $x$
by letting $a \in A$, and considering the map $(x,P) \mapsto
M_{\tau_x(a)}(P)$. The same holds for $p,q,y$. It follows that the
poles of $M_{\tau_x(a)}(P)$ depend on $x$, hence we obtain the
``moving poles" $P_1(x),\ldots,P_n(x)$. In addition, it is clear
that $P_1(0)=P_1,\ldots,P_n(0)=P_n$ are the finite poles of the
function
$M_a(P)$.\\

Based on this algebro-geometric structure, we are able to find
formulas for the functions $p(x), q(x), y(x)$ in terms of the finite
poles and the zeros of $M_{\tau_x(a)}(P)$, for every $a \in A$. For
our convenience, we write $P_i(x)$ instead of $\pi(P_i(x))$, if no
confusion arises.

\newtheorem{the2.5}[defi2.1]{Theorem}
 \begin{the2.5}
 Suppose that Hypotheses $2.4$ hold. Then for every $a \in A$, the
 finite poles $P_j(x)$ $(j=1,\ldots,n)$ of the function
 $M_{\tau_x(a)}(P)$ satisfy the following first-order system of
 differential equations
    \begin{equation}\label{2.3}
     P_{j,x}(x)=
 \frac{(-1)^n k(P_j(x))\sqrt{P_j(x)} ~[m_-^0(x)-m_+^0(x)] \prod_{i=1}^n P_i(x)}
      {p(x) k(0) \prod_{s \neq j} (P_j(x)-P_s(x))},
    \end{equation}
where $m_{\pm}^0(x):=m_\pm(\tau_x(a),0).$
 Moreover, the functions $p(x), q(x), y(x)$
 satisfy the following relations involving the finite poles of the
 meromorphic function $M_{\tau_x(a)}(P)$ on $\mathcal{R}$,
  \begin{equation*}
  \frac{2\sqrt{py}}{p}=\frac{T_x(P_j(x))}{k(P_j(x)) \sqrt{P_j(x)}}
  =\frac{(-1)^{n+1}[m_-^0(x)-m_+^0(x)]
   \prod_{i=1}^n P_i(x)}{p(x)k(0)},
  \end{equation*}
and
  \begin{equation*}
   q(x)=y(x) \Big( \sum_{j=1}^{2n} z_j -2\sum_{i=1}^n P_i(x)
   \Big)
   +q_{n,x}(x)+\frac{q_n^2(x)}{p(x)},
  \end{equation*}
where
  \begin{eqnarray*}
   q_n(x)&=&\frac{i (p y)^\prime}{4y}=\frac{i p(x)}{2} \frac{d}{dx}
   \left( \mathrm{ln} \Big([m_-^0(x)-m_+^0(x)] \prod_{i=1}^n P_i(x) \Big)
   \right)
     \\
   &=&
   \sum_{j=1}^n \frac{\sqrt{p(x) y(x)} k(P_j(x))}
   {\sqrt{P_j(x)} \displaystyle \prod_{r \neq j} (P_j(x)-P_r(x)) }
   -\sum_{j=1}^n \frac{Q(0,x)}{\sqrt{P_j(x)} \displaystyle \prod_{r \neq j}
   (P_j(x)-P_r(x))}.
  \end{eqnarray*}
 \end{the2.5}
The proof of this theorem can be found in \cite{38}.

What is of great relevance is the fact that this problem can be
inverted, i.e. the following holds.

\newtheorem{the2.6}[defi2.1]{Theorem}
 \begin{the2.6}
 Let $z_1<z_2<\ldots<z_{2n}$ be distinct positive real numbers, and
 let $\mathcal{R}$ be the Riemann surface genus of $n$ described
 by the following algebraic relation
   \begin{equation*}
   w^2=z(z-z_1)(z-z_2)\cdots(z-z_{2n}).
   \end{equation*}
 Let $\tilde{P}_1,\ldots,\tilde{P}_n$ be points on $\mathcal{R}$
 such that $z_{2j-1} \leq \tilde{P}_j \leq z_{2j}$ $(1\leq j \leq
 n)$. Then there exists a stationary ergodic process $(A,
 \{\tau_x\}, \mu)$ together with functions $p,q,y: A \rightarrow
 \mathbb{R}$, such that the spectrum of the family
   \begin{equation}
\left(
  \begin{array}{c}
    v_1 \\
    v_2 \\
  \end{array}
\right)^\prime=
 \left(
   \begin{array}{cc}
     0 & 1/p(\tau_x(a)) \\
     q(\tau_x(a))-zy(\tau_x(a)) & 0 \\
   \end{array}
 \right)
\left(
  \begin{array}{c}
    v_1 \\
    v_2 \\
  \end{array}
\right), \quad a \in A, z\in \mathbb{C},
   \end{equation}
of differential equations does not depend on the choice of $a \in A$
and has the form
  \begin{equation}
  \Sigma=[z_2,z_3] \cup [z_4,z_5] \cup \dots \cup [z_{2n},\infty).
  \end{equation}
Moreover, $\beta(z)=0$, for Lebesgue a.e. $z\in \Sigma$. The
functions $p,q,y$ satisfy relations analogous to those of Theorem
$2.5$.
\end{the2.6}
The proof of this theorem can be found in \cite{38}. The importance
of Theorem 2.6 lies in the fact that we can generate infinitely many
stationary ergodic processes and functions $p,q,y$ such that the
above conclusions hold. For more explanations about this importance,
see again \cite{38}.

\section{The stationary HS hierarchy}
In this section, we derive the stationary HS hierarchy and the
corresponding sequence of zero-curvature pairs by using a polynomial
recursion formalism. We will use the ergodic-dynamical structure of
Section 2 in the following.

Let $\mathcal{L}$ be the set of all the positive uniformly
continuous bounded functions $f: \mathbb{R} \rightarrow \mathbb{R}.$
We equip $\mathcal{L}$ with the compact open topology. Let $\{
\tau_x \}$ be the \textit{Bebutov flow} on $\mathcal{L}$, that is,
$\tau_x(f)(\cdot)=f(x+\cdot),$ for every $f \in \mathcal{L}.$
Moreover, we define $\mathcal {A}=Hull(y_0)=cls\{\tau_x(y_0), ~
x\in\mathbb{R},~ y_0\in\mathcal{L}\}.$ One infers that $\mathcal{A}$
is compact. For every $A \in \mathcal{A}$, let $y(A)=A(0)$. We
consider the family of equations
  \begin{equation}\label{3.1}
   X'=\left(
        \begin{array}{c}
          \psi_1 \\
          \psi_2 \\
        \end{array}
      \right)'
      =\left(
         \begin{array}{cc}
           0 & 1 \\
           -zy(\tau_x(A)) & 0\\
         \end{array}
       \right)
       \left(
        \begin{array}{c}
          \psi_1 \\
          \psi_2 \\
        \end{array}
      \right)
      =A(x)X,
      \quad A\in \mathcal{A},~z\in\mathbb{C},
  \end{equation}
which is equivalent to $-\psi_1''=z y \psi_1.$

Now we start with the following  $2\times 2$ matrix isospectral
problem
     \begin{equation}\label{3.2}
       \psi_x=A(x,z)\psi
     \end{equation}
and an auxiliary problem
    \begin{equation} \label{3.3}
       \psi_{t_n}=B_n(z)\psi,
    \end{equation}
where $A(x,z)$ is a Sturm-Liouville matrix, defined by
    \begin{equation}\label{3.4}
     A(x,z)=
       \left(
         \begin{array}{cc}
           0 & 1 \\
           -zy(x) & 0 \\
         \end{array}
       \right)
     \end{equation}
and  $B_n(z)$ is a two dimensional matrix-valued linear differential
operator, defined by
    \begin{equation}\label{3.5}
      B_n(z)=
      \left(
        \begin{array}{cc}
          -G_n(z) & z^{-1}F_{n}(z) \\
          -H_n(z) & G_n(z) \\
        \end{array}
      \right)
    \qquad z \in \mathbb{C} \setminus \{0\},
    \quad n\in\mathbb{N},
    \end{equation}
assuming $F_{n}$, $G_n$ and $H_n$ to be polynomials of degree at
most $n$ with $C^\infty$ coefficients with respect to $x$. The
compatibility condition between  (\ref{3.2}) and (\ref{3.3}) yields
the stationary zero-curvature equation
\begin{equation}\label{3.6}
   -B_{n,x}+[A,B_n]=0,
\end{equation}
namely
   \begin{eqnarray}\label{3.7}
      F_{n,x}&=& 2zG_n, \\
      H_{n,x}&=& -2zyG_n, \\
      G_{n,x}&=&H_n -yF_{n}.
    \end{eqnarray}
From (\ref{3.7})-(3.9), a direct calculation shows that
   \begin{equation}\label{3.10}
     \frac{d}{dx} \mathrm{det} (B_n(z,x))=
     -\frac{1}{z} \frac{d}{dx} \Big(
     zG_n(z,x)^2-F_{n}(z,x)H_n(z,x)
     \Big)=0.
   \end{equation}
Hence, $zG_n^2-F_{n}H_n$ is $x$-independent implying
   \begin{equation}\label{3.11}
    zG_n^2-F_{n}H_n=k^2(z),
   \end{equation}
where the integration constant $k^2(z)$ is a polynomial of degree
$2n$ with respect to $z$. If $\{z_m\}_{m=1,\cdots,2n}$ denote its
zeros, then
   \begin{equation}\label{3.12}
   k^2(z)=\prod_{m=1}^{2n}(z-z_m),\quad
   \{z_m\}_{m=1,\ldots,2n}\in\mathbb{C}.
   \end{equation}

Next, we introduce the following polynomial $F_{n}(z), G_n(z)$ and
$H_n(z)$ with respect to the spectral parameter $z$,
   \begin{equation}\label{3.13}
     F_{n}(z)=\sum_{l=0}^{n} f_{l} z^{l},
   \end{equation}

   \begin{equation}\label{3.14}
     G_n(z)=\sum_{l=0}^{n-1} g_{l} z^{l},
   \end{equation}

   \begin{equation}\label{3.15}
     H_n(z)=\sum_{l=0}^n h_{l} z^{l}.
   \end{equation}
From (\ref{3.7}), (3.8) and (\ref{3.11}), we obtain that $f_0, h_0$
is a constant and $f_0h_0=-\prod z_m$. For our convenience in the
following, we choose $f_0$ such that
    \begin{equation}\label{3.16}
     f_0=-h_0=-\sqrt{\prod z_m}.
    \end{equation}
Without loss of generality, let $f_0=h_0=1$ to normalize the
polynomials $F_n$ and $H_n$. We keep the same notation for the
normalized polynomials, unless explicitly stated.

We begin our manipulation.

First, substituting (\ref{3.7}) into (3.9) and (3.8), we arrive at
   \begin{equation}\label{3.17}
    \frac{1}{2}F_{n,xx}-zH_n+zyF_n=0,
   \end{equation}
   \begin{equation}\label{3.18}
   H_{n,x}+yF_{n,x}=0.
   \end{equation}
Then, differentiating (\ref{3.17}) with respect to $x$, we obtain
   \begin{equation}\label{3.19}
    \frac{1}{2}F_{n,xxx}=zH_{n,x}-zy_xF_n-zyF_{n,x}.
   \end{equation}
By subtraction and addition of (\ref{3.19}) to (\ref{3.18}), we have
    \begin{equation}\label{3.20}
     \frac{1}{2}F_{n,xxx}=-2zyF_{n,x}-zy_xF_n,
    \end{equation}
    \begin{equation}\label{3.21}
    \frac{1}{2}F_{n,xxx}=2zH_{n,x}-zy_xF_n.
    \end{equation}
Hence, comparing the coefficients of the same powers in
(\ref{3.20}),  we have the following recursion formalism for the
coefficients $f_l$:
    \begin{equation}\label{3.22}
     f_{l,x}=-\mathcal{G}(4yf_{l-1,x}+2y_xf_{l-1}),
    \end{equation}
where $\mathcal{G}$ is given by
    \begin{equation}\label{3.23}
     (\mathcal{G}v)(x)=\int_{-\infty}^x \int_{-\infty}^{x_1}
     v(x_2) ~dx_2 dx_1 , \quad x\in\mathbb{R},~
        v\in L^\infty(\mathbb{R}).
     \end{equation}
It is easy to see that $\mathcal{G}$ is the resolvent of the
one-dimensional Laplacian operator, that is
    \begin{equation}\label{3.24}
        \mathcal{G}=\Big( \frac{d^2}{dx^2} \Big)^{-1}.
    \end{equation}
Explicitly, we compute
    \begin{equation}\label{3.25}
      \begin{split}
       & f_0=1, \\
       & f_1=-u+c_1,\\
       & f_2=\mathcal{G}(u_{xx}u+\frac{1}{2}u_x^2)-uc_1+c_2, \\
       & \textrm{etc}.
      \end{split}
    \end{equation}
where $\{c_l\}_{l\in\mathbb{N}}\subset\mathbb{C}$ are integration
constants and we have used the assumption $f_l(u)|_{u=0}=c_l$, $l
\in \mathbb{N}$.

The relation (\ref{3.7}) and (\ref{3.21}) provide the recursion
formalism for the coefficients $g_l$ and $h_l$ of the polynomial
$G_n$ and $H_n$, one can refer to \cite{16}.

For fixed $n \in \mathbb{N}$, by using (3.9), for $l=n$ we obtain
$h_n=yf_n$, and together with (\ref{3.21}) yields,
       \begin{equation}\label{3.26}
        \textrm{s-HS}_n(u)=2u_{xx}f_{n,x}(u)+u_{xxx}f_n(u)=0.
       \end{equation}
This is the stationary HS equation of order $n$.

Hence, we obtain the stationary HS hierarchy. The first equation in
the hierarchy, obtained by taking $n=1$, we compute explicitly,
       \begin{equation}\label{3.27}
        \textrm{s-HS}_1(u)=-2u_{xx}u_x-u_{xxx}u+u_{xxx}c_1=0.
       \end{equation}
It is easy to see that $\textrm{s-HS}_1(u)=0$ ($c_1=0$) represents
the classical one dimensional stationary HS equation.

\section{The time-dependent HS hierarchy}
In this section,  we will introduce the time-dependent HS hierarchy.
This means that $u$ are now considered as functions of both space
and time. Let $r\in\mathbb{N}$ be fixed. We introduce a deformation
parameter $t_r \in \mathbb{R}$ in $u$, replacing $u(x)$ by
$u(x,t_r)$, for each equation in the hierarchy.

Now we consider the Sturm-Liouville matrix
   \begin{equation}
    A=\left(
        \begin{array}{cc}
          0 & 1\\
          -zy(x,t_r) & 0 \\
        \end{array}
      \right)
   \end{equation}
and the matrix $B_r(x,t_r)$ whose entries are polynomials $F_r$ and
$H_r$ of degree $r$ in $z$, $G_r$ of degree $r-1$ in $z$, and with
coefficients depending on $t_r$ as well. Then the compatibility
condition yields the zero-curvature equation
   \begin{equation}\label{4.2}
   A_{t_r}-B_{r,x}+[A,B_r]=0, \qquad r\in \mathbb{N},
   \end{equation}
namely
   \begin{eqnarray}\label{4.3}
      && -zy_{t_r}+H_{r,x}+2zyG_r=0, \\
      && F_{r,x}=2zG_r, \\
      && G_{r,x}=H_r -yF_{r}.
    \end{eqnarray}
Substituting (4.4) into (4.5) and (4.3), we get
   \begin{equation}\label{4.6}
    \frac{1}{2}F_{r,xx}=zH_r-zyF_r,
   \end{equation}
   \begin{equation}\label{4.7}
    -zy_{t_r}+H_{r,x}+yF_{r,x}=0.
   \end{equation}
Then as in the stationary case, differentiating (\ref{4.6}) with
respect to $x$, and by addition and subtraction to (\ref{4.7}), we
arrive at
  \begin{equation}\label{4.8}
  \frac{1}{2}F_{r,xxx}=z^2y_{t_r}-2zyF_{r,x}-zy_xF_r,
  \end{equation}
  \begin{equation}\label{4.9}
   \frac{1}{2}F_{r,xxx}=-z^2y_{t_r}+2zH_{r,x}-zy_xF_r.
  \end{equation}
Hence, the relation (\ref{4.8}) gives the recursion formalism for
the coefficients $f_l$:
   \begin{equation}\label{4.10}
    \begin{split}
    & f_1=-u+c_1, \\
    & \frac{1}{2}f_{2,xxx}=y_{t_r}-2yf_{1,x}-y_xf_1,\\
    &
    f_{l,x}=-\mathcal{G}(4yf_{l-1,x}+2y_xf_{l-1}-2\delta_{l,2}y_{t_r}).
    \end{split}
   \end{equation}
From (4.10), we infer that all the higher order coefficients depend
in an implicit way on $t_r$, since $f_2$ depends on $t_r$.

Moreover, the coefficients $h_l$ of $H_r$ are determined by
(\ref{4.9}). Since from (4.5), we obtain $h_r=yf_r$, then together
with (\ref{4.9}) yields
  \begin{equation}\label{4.11}
   \textrm{HS}_r(u)=2y(x,t_r)f_{r,x}(x,t_r)+y_x(x,t_r)f_r(x,t_r)=0,
   \quad r \geq 2.
  \end{equation}
which is the $r$-th order HS equation. In addition, in the case
$r=1$, the corresponding equation can also be derived from
(\ref{4.9}), that is,
   \begin{equation}\label{4.12}
     \textrm{HS}_1(u)=u_{xxt_1}+2u_{xx}u_x+uu_{xxx}-u_{xxx}c_1=0.
   \end{equation}
It is clear that $\textrm{HS}_1(u)=0$ ($c_1=0$) represents the
classical HS equation.

\section{The stationary HS formalism}
In this section we focus our attention on the stationary case. By
solving the inverse Sturm-Liouville problem, we obtain the relation
between the Weyl $m$-functions $m_\pm$ and the entries of the matrix
$B_n$ of the stationary HS hierarchy, which will bring to the
determination of the solution of the HS hierarchy both in stationary
and time-dependent cases.\\

Consider the polynomials $F_n, G_n$ and $ H_n$ as defined before in
the stationary case, then (\ref{3.11}) gives,
   \begin{equation}\label{5.1}
    zG_n^2-F_nH_n=k^2(z)=\prod_{m=1}^{2n}(z-z_m).
   \end{equation}
We introduce the hyperelliptic curve $\mathcal{K}_n$,
   \begin{equation}\label{5.2}
    \mathcal{K}_n: w^2-zk^2(z)=w^2-z\prod_{m=1}^{2n}(z-z_m)=0, \quad
      \{z_m\}_{m=1,\ldots, 2n} \in \mathbb{C},
    \end{equation}
which is compactified by joining the point $P_{\infty}$ at infinity.
The complex structure on $\mathcal{K}_{n}$ is defined in the usual
way \cite{12}. Hence, $\mathcal{K}_{n}$ becomes a two-sheeted
hyperelliptic Riemann surface of genus $n$ (possibly with a singular
affine part) in a standard manner, denoted by $\mathcal{R}$.

Let $\mu_j,\nu_j: \mathbb{R} \rightarrow \mathbb{R}$
$(j=1,\ldots,n)$ be continuous functions with $\mu_i \neq \mu_j$,
$\nu_i \neq \nu_j$ for $ i \neq j$, and such that
   \begin{equation}\label{5.3}
    F_n(z)=\frac{(-1)^{n+1} k(0)}{\prod_{j=1}^n \mu_j(x)}
          \prod_{j=1}^n (z-\mu_j(x))
   \end{equation}
and
   \begin{equation}\label{5.4}
    H_n(z)=\frac{(-1)^{n} k(0)}{\prod_{j=1}^n \nu_j(x)}
          \prod_{j=1}^n (z-\nu_j(x)).
   \end{equation}

Next, we define the function $\tilde{M}: \mathcal{R} \times
\mathbb{R} \rightarrow \mathbb{C}$,
   \begin{equation}\label{5.5}
    \tilde{M}(P,x)=\frac{k(P) + \sqrt{z} G_n(z)}{F_n(z)}
    =\frac{-H_n(z)}{k(P)-\sqrt{z}G_n(z)},
    \quad \pi(P)=z.
   \end{equation}

\newtheorem{the5.1}{Theorem}[section]
 \begin{the5.1}
  For every fixed  $P \in \mathcal{R}$ such that $\pi(P)=z \in
  \mathbb{C} \setminus \Sigma$, $\tilde{M}(P,x)$ satisfies the
  following Riccati equation
     \begin{equation}\label{5.6}
       \tilde{M}_x(P,x)+z^{1/2}\tilde{M}(P,x)^2=-z^{1/2}y.
     \end{equation}
 \end{the5.1}
\textbf{Proof.}~~ Using (\ref{5.5}), (3.7), (3.9) and (\ref{3.11}),
a direct calculation shows that (\ref{5.6}) is true. \quad
$\square$ \\

The following theorem describes the nature of the $x$-motion of the
zeros $\{\mu_j(x)\}_{j=1,\ldots,n}$ of $F_n$.

\newtheorem{the5.2}[the5.1]{Theorem}
  \begin{the5.2}
   Assume that the zeros $\{\mu_j(x)\}_{j=1,\ldots,n}$ of $F_n$ remain
   distinct, then $\{\mu_j(x)\}_{j=1,\ldots,n}$ satisfy the system of
   differential equations,
     \begin{equation}\label{5.7}
      \mu_{j,x}(x)=
      \frac{2 (-1)^n  k(\mu_j(x)) \sqrt{\mu_j(x)} \prod_{i=1}^n \mu_i(x)}
       {k(0) \prod_{\scriptstyle i=1 \atop \scriptstyle i \neq j}^n
       (\mu_j(x)-\mu_i(x))}.
     \end{equation}
   Moreover, given initial data satisfying $\mu_j(x_0) \in
   [z_{2j-1},z_{2j}]$ $(j=1,\ldots,n),$ then
      \begin{equation}\label{5.8}
       \mu_j(x) \in  [z_{2j-1},z_{2j}], \quad
       j=1,\ldots,n, \quad x\in\mathbb{R}.
      \end{equation}
   In particular, $\mu_j(x)$ changes sheets whenever it hits $z_{2j-1}$
   or $z_{2j}$.
  \end{the5.2}
\textbf{Proof.}~~The derivatives of (\ref{5.3}) with respect to $x$
take on
     \begin{equation}\label{5.9}
      F_{n,x}(\mu_j)=\frac{(-1)^n k(0) }{\prod \mu_j} \mu_{j,x}
      \prod_{\scriptstyle i=1 \atop \scriptstyle i \neq j}^n (\mu_j-\mu_i).
     \end{equation}
On the other hand, inserting $z=\mu_j$ into (\ref{3.7}) yields
      \begin{equation}\label{5.10}
       F_{n,x}(\mu_j)=2\mu_jG_n(\mu_j)=2\sqrt{\mu_j}k(\mu_j).
      \end{equation}
Comparing (\ref{5.9}) and (\ref{5.10}) leads to (\ref{5.7}). To see
(\ref{5.8}), one can use the local coordinate $
\lambda=(z-z_\ast)^{1/2}$ near a ramification point $z_\ast$
\cite{12,16}.
\quad $\square$ \\

Now we explain the importance of the relation (\ref{5.7}) from two
aspects. First, it shows the behavior of a point $\mu_j(x)$ when it
reaches a ramification point of the resolvent interval containing
it. In fact, as soon as the point $\mu_j(x)$ reaches a ramification
point $z_\ast$, its derivative becomes zero ($k(z_\ast)=0$), and it
passes from one sheet to the other in the Riemann surface
$\mathcal{R}$, i.e, $\mu_j(x) \in [z_{2j-1},z_{2j}]$ for every
$j=1,\ldots,n$. Second, from Theorem 2.6, we know that for every
choice of points $\tilde{P}_j \in [z_{2j-1},z_{2j}]$
($j=1,\ldots,n$), there exists a meromorphic function $P \rightarrow
M(x,P)$ on $\mathcal{R}$, having as finite simple poles the points
$P_j(x) $ $(j=1,\ldots,n)$, satisfying (\ref{2.3}), which, adapted
to the present setting ($p=1$, $q=0$, $m_-^0(x)=-m_+^0(x)=1$)
becomes
     \begin{equation}\label{5.11}
      P_{j,x}(x)=\frac{2 (-1)^n k(P_j(x)) \sqrt{P_j(x)} \prod_{i=1}^n
      P_i(x)}
      {k(0) \prod_{s \neq j} (P_j(x)-P_s(x))}.
     \end{equation}
Hence, the points $\mu_j(x)$ satisfy the system (\ref{5.11}), as
well as the poles of the meromorphic function $M_y$ on $\mathcal{R}$
defined as in (\ref{2.2}).

At this point we can use Theorem 2.6 again to prove that there is a
function $y(x)$ satisfying the assumptions on the spectrum (H1) and
(H2), and such that the poles $P_j(x)$ of the function $M_y$ satisfy
$\pi(P_j(x))=\mu_j(x)$, for every $j=1,\ldots,n$.

Another important fact from (\ref{5.7}) is that $\mu_j(x) \in
C^\infty (\mathbb{R})$, which has been proved in our latest paper
\cite{16}.

Based on such a $y(x)$, obtained by solving the inverse
Sturm-Liouville problem, we obtain the following fundamental
theorem.

\newtheorem{the5.3}[the5.1]{Theorem}
 \begin{the5.3}
  Let $P_j(x)$ $(j=1,\ldots,n)$ be $n$ distinct points on the
  Riemann surface $\mathcal{R}$ satisfying $(\ref{5.11})$, and such
  that $P_j(x_0)=\tilde{P}_j \in [z_{2j-1},z_{2j}]$
  $(j=1,\ldots,n)$. Then there exists a positive continuous function
  $y(x)$ such that
     \begin{equation}\label{5.12}
      F_n(z)=\frac{1}{\sqrt{y(x)}}
      \prod_{j=1}^n (z- \pi(P_j(x))),
     \end{equation}
     \begin{equation}\label{5.13}
      M_y(P,x)=\tilde{M}(P,x)=m_+(P,x)
      \quad
  \textrm{and} \quad
   \tilde{M}(P,x) \circ \sigma (P)=m_-(P,x),
     \end{equation}
  for every $x\in\mathbb{R}$ and $P\in \mathcal{R}$.
  Moreover, with such $y(x)$, Hypotheses $2.4$ hold.
 \end{the5.3}

\newtheorem{rem5.4}[the5.1]{Remark}
 \begin{rem5.4}
  The similar formulas describing the nature of the $x$-motion of
  the points $\nu_j(x)$ can be found in \cite{16}, here
  we omit the details. In addition, the regularity properties,
  together with the other observations concerning the motion which
  are valid for $\mu_j(x)$,
  also hold for $\nu_j(x)$.
 \end{rem5.4}

We must emphasize an important difference between our approach in
\cite{16} and that in the present paper: while in the former
approach, we deduce the algebro-geometric structure directly from
the hierarchy, in the latter we start from an assigned
algebro-geometric structure and prove that it is conserved by the
hierarchy.

Given these preparations, one of the main results of this section,
that is, the algebro-geometric formula for the solution $u(x)$ of
the stationary HS hierarchy reads as follows.

\newtheorem{the5.5}[the5.1]{Theorem}
 \begin{the5.5}
  The solution $u(x)$ of the stationary HS hierarchy is
    \begin{equation}\label{5.14}
     u(x)=\sum_{j=1}^n
     \frac{1}{P_j(x)}-\sum_{m=1}^{2n}\frac{1}{z_m},
    \end{equation}
where the points $P_j(x)$ $(j=1,\ldots,n)$ evolve according to
$(\ref{5.11})$.
\end{the5.5}
\textbf{Proof.}~~Let $z=0$ in (\ref{5.5}), we obtain
   \begin{equation}\label{5.15}
    m_+^0(x)=-1,
    \quad
    m_-^0(x)=+1,
   \end{equation}
where we used (\ref{3.16}) and (\ref{5.13}). Hence, take $p=1$ in
Theorem 2.5, we arrive at
   \begin{equation}\label{5.16}
    \sqrt{y(x)}=\frac{(-1)^{n+1} }{k(0)} \prod_{j=1}^n P_j(x).
   \end{equation}
Next, we consider the coefficient of $z$ in (\ref{5.12}),
   \begin{equation}\label{5.17}
    f_1(x)=\frac{(-1)^{n-1}}{\sqrt{y(x)}}
    \Big( \sum_{j=1}^n \prod_{\scriptstyle i=1 \atop \scriptstyle i
    \neq j}^n P_i(x)
    \Big).
   \end{equation}
Inserting (\ref{5.16}) into (\ref{5.17}), we get
   \begin{equation}\label{5.18}
   f_1(x)=\sqrt{\prod_{m=1}^{2n} z_m}
          \Big( \sum_{j=1}^n \frac{1}{P_j(x)}
          \Big).
   \end{equation}
On the other hand, we note that the coefficient $f_1(x)$ in
(\ref{3.25}) is not normalized, we find
    \begin{equation}\label{5.19}
      u(x)-c_1=\frac{1}{\sqrt{\prod_{m=1}^{2n} z_m}}f_1(x).
    \end{equation}
Hence
    \begin{equation}\label{5.20}
    u(x)=\sum_{j=1}^n \frac{1}{P_j(x)} +c_1.
    \end{equation}
The constant $c_1$ can be determined from (\ref{3.11}), by computing
the coefficient of $z$, we obtain
  \begin{equation}\label{5.21}
   c_1=-\sum_{m=1}^{2n}\frac{1}{z_m}.
  \end{equation}

\section{The time-dependent HS formalism}
The basic problem in an algebro-geometric construction of the
solutions of the time-dependent HS hierarchy is to solve the
time-dependent $r$-th HS equation with a stationary solution of the
$n$-th equation as initial data in the hierarchy (recall that
$r<n$).

We employ the notations $\widetilde{B}_r,$ $\widetilde{F}_{r},$
$\widetilde{G}_r,$ $\widetilde{H}_r,$ $\tilde{f}_{r}$,
$\tilde{g}_{r},$ $\tilde{h}_{r}$ to stand for the time-dependent
quantities, and we will keep the usual notations for the stationary
quantities.

Summing up, we are seeking a solution $u(x,t_r)$ of the
time-dependent algebro-geometric initial value problem
\begin{eqnarray}\label{6.1}
    &&
     \begin{split}
       & \mathrm{HS}_r(u)=2y(x,t_r)\tilde{f}_{r,x}(x,t_r)
          +y_x(x,t_r)\tilde{f}_{r}(x,t_r)=0, \\
       & u|_{t_r=t_{0,r}}=u_0,
      \end{split}
          \\
   &&
      \textrm{s-HS}_n(u_0)=u_{0,xxx}f_{n}
      +2u_{0,xx}f_{n,x}=0,
   \end{eqnarray}
where
   \begin{equation}\label{6.3}
    2y(x,t_r)=u_{xx}(x,t_r).
   \end{equation}

We start from the zero-curvature equations:
    \begin{equation}\label{6.4}
        A_{t_r}(x,t_r)-\widetilde{B}_{r,x}(x,t_r)
        +[A(x,t_r),\widetilde{B}_r(x,t_r)]=0,
    \end{equation}
    \begin{equation}\label{6.5}
        -B_{n,x}(x,t_r)+[A(x,t_r),B_n(x,t_r)]=0,
    \end{equation}
where
   \begin{equation}\label{6.6}
    \widetilde{B}_r(z)=
       \left(
         \begin{array}{cc}
           -\widetilde{G}_r(z) & z^{-1}\widetilde{F}_{r}(z) \\
           -\widetilde{H}_r(z) & \widetilde{G}_r(z) \\
         \end{array}
       \right)
   \end{equation}
with entries
   \begin{eqnarray}\label{6.7}
 &&
    \widetilde{F}_{r}(z)=\sum_{s=0}^{r} \tilde{f}_{s}(x,t_r) z^{s},
      \\
   &&
    \widetilde{G}_r(z)=\sum_{s=0}^{r-1} \tilde{g}_{s}(x,t_r) z^{s},
    \\
    &&
    \widetilde{H}_r(z)=\sum_{s=0}^r \tilde{h}_{s}(x,t_r) z^{s}.
   \end{eqnarray}

Moreover, it is more convenient for us to rewrite the zero-curvature
equations (\ref{6.4}) and (\ref{6.5}) as the following forms,
   \begin{equation}\label{6.10}
        -zy_{t_r}+\widetilde{H}_{r,x}
        +2zy\widetilde{G}_r=0,
    \end{equation}
     \begin{equation}\label{6.11}
        \widetilde{F}_{r,x}
        =2z\widetilde{G}_r,
    \end{equation}
    \begin{equation}\label{6.12}
        \widetilde{G}_{r,x}=\widetilde{H}_r
         -y\widetilde{F}_{r},
    \end{equation}
and
   \begin{equation}\label{6.13}
     F_{n,x}=2zG_n,
   \end{equation}
   \begin{equation}\label{6.14}
     H_{n,x}=-2zyG_n,
   \end{equation}
   \begin{equation}\label{6.15}
        G_{n,x}=H_n-yF_{n}.
    \end{equation}

From (\ref{6.13})-(\ref{6.15}), we may compute
   \begin{equation}\label{6.16}
    \frac{d}{dx} \mathrm{det}(B_n(z))=-\frac{1}{z}\frac{d}{dx}
    \Big( zG_n(z)^2-F_{n}(z)H_n(z) \Big)=0,
   \end{equation}
and meanwhile Lemma 6.3 gives
  \begin{equation}\label{6.17}
    \frac{d}{dt_r} \mathrm{det}(B_n(z))=-\frac{1}{z}\frac{d}{dt_r}
    \Big( zG_n(z)^2-F_{n}(z)H_n(z) \Big)=0.
  \end{equation}
Hence, $zG_n(z)^2-F_{n}(z)H_n(z)$ is independent of variables both
$x$ and $t_r$, which implies
   \begin{equation}\label{6.18}
   zG_n(z)^2-F_{n}(z)H_n(z)=k^2(z).
   \end{equation}
This reveals that the fundamental identity (\ref{3.11}) still holds
in the time-dependent context. Consequently the hyperelliptic curve
$\mathcal{K}_n$ is still available by (\ref{5.2}).

Next, we define
   \begin{equation}\label{6.19}
    m_+(P,x,t_r)=\frac{k(P) + \sqrt{z}G_n(z) }{F_n(z)},
    \quad \pi(P)=z
    \end{equation}
and
  \begin{equation}\label{6.20}
   m_-(P,x,t_r)=\frac{-k(P) + \sqrt{z}G_n(z) }{F_n(z)},
   \quad \pi(P)=z.
  \end{equation}

The properties of the Weyl $m$-functions $m_\pm(P,x,t_r)$ are
summarized as follows.

\newtheorem{lem6.1}{Lemma}[section]
 \begin{lem6.1}
  The Weyl $m$-functions $m_\pm(P,x,t_r)$ satisfy the following
  Riccati equation,
   \begin{equation}\label{6.21}
    M_x(P,x,t_r)+z^{1/2}M(P,x,t_r)^2=-z^{1/2}y(x,t_r).
    \end{equation}
 Moreover,
  \begin{equation}\label{6.22}
   m_+(P,x,t_r)+m_-(P,x,t_r)=\frac{2 \sqrt{z} G_n(z)}{F_n(z)},
  \end{equation}
  \begin{equation}\label{6.23}
   m_+(P,x,t_r)-m_-(P,x,t_r)=\frac{2k(P)}{F_n(z)},
  \end{equation}
  \begin{equation}\label{6.24}
   m_+(P,x,t_r)m_-(P,x,t_r)=\frac{H_n(z)}{F_n(z)}.
  \end{equation}
 \end{lem6.1}
\textbf{Proof.}~~The proof of (\ref{6.21}) is identical to
(\ref{5.6}). The relations (\ref{6.22})-(\ref{6.24}) are an
immediate consequence of the definitions of $m_\pm$ and the
fundamental identity (\ref{6.18}). \quad $\square$

\newtheorem{lem6.2}[lem6.1]{Lemma}
 \begin{lem6.2}
  Assume that $(\ref{6.4})$ and $(\ref{6.5})$ hold. Let
  $P \in \mathcal{R} \setminus \{P_{\infty}\}$ and $(x,t_r) \in
  \mathbb{R}^2$. Then the function $M(P,x,t_r)$ satisfies the
  following differential equations
   \begin{equation}\label{6.25}
   M_{t_r}(P,x,t_r)=-z^{-\frac{1}{2}}\widetilde{H}_r(z)
   +2\widetilde{G}_r(z)
   M(P,x,t_r)-z^{-\frac{1}{2}}\widetilde{F}_r(z)
   M(P,x,t_r)^2,
   \end{equation}
  and
  \begin{eqnarray}\label{6.26}
   \begin{split}
  M_{t_r}(P,x,t_r)&=(-z^{-\frac{1}{2}}\widetilde{G}_r(z)+z^{-1}\widetilde{F}_r(z)
  M(P,x,t_r))_x \\
  &=-z^{-\frac{1}{2}}\widetilde{H}_r(z)+z^{-\frac{1}{2}}y\widetilde{F}_r(z)
  +z^{-1}(\widetilde{F}_r(z)M(P,x,t_r))_x.
  \end{split}
  \end{eqnarray}
 \end{lem6.2}
\textbf{Proof.}~~Using (\ref{6.10})-(\ref{6.12}) and (\ref{6.21}),
by a straightforward but rather lengthy calculation, we infer
   \begin{equation}\label{6.27}
    (\partial_x + 2z^{\frac{1}{2}}M)
    (M_{t_r}+ z^{-\frac{1}{2}}\widetilde{H}_r-2\widetilde{G}_r M
    +z^{-\frac{1}{2}}\widetilde{F}_r M^2)=0.
   \end{equation}
Hence
   \begin{equation}\label{6.28}
    M_{t_r}+ z^{-\frac{1}{2}}\widetilde{H}_r-2\widetilde{G}_r M
    +z^{-\frac{1}{2}}\widetilde{F}_r M^2
    =C ~ \mathrm{exp} \Big( -2 \int^x
    z^{\frac{1}{2}}M ~dx^\prime
    \Big),
   \end{equation}
where the left-hand side is meromorphic in a neighborhood of
$P_{\infty}$, while the right-hand side is meromorphic near
$P_{\infty}$ only if $C=0$. This proves (\ref{6.25}). Next, by using
(\ref{6.11}) and (\ref{6.21}), we obtain
  \begin{eqnarray}\label{6.29}
   z^{-\frac{1}{2}} y \widetilde{F}_r +z^{-1}(\widetilde{F}_r M)_x
   &=&z^{-\frac{1}{2}} y \widetilde{F}_r+z^{-1}\widetilde{F}_{r,x}M
   +z^{-1}\widetilde{F}_r M_x
   \nonumber \\
   &=&
   2\widetilde{G}_r M-z^{-\frac{1}{2}} \widetilde{F}_r M^2.
     \end{eqnarray}
Combining this result with (\ref{6.25}), we conclude that
(\ref{6.26}) holds. Alternatively, more efficiently method to prove
(\ref{6.25}) and (\ref{6.26}) can be found in our paper \cite{16}.
\quad $\square$ \\

Next, we study the time evolution of $F_n$, $G_n$ and $H_n$ by using
zero-curvature equations (\ref{6.10})-(\ref{6.12}) and
(\ref{6.13})-(\ref{6.15}).

\newtheorem{lem6.3}[lem6.1]{Lemma}
 \begin{lem6.3}
  Assume that $(\ref{6.4})$ and $(\ref{6.5})$ hold. Then
  \begin{equation}\label{6.30}
    F_{n,t_r}=2(G_n\widetilde{F}_{r}-\widetilde{G}_rF_{n}),
   \end{equation}
   \begin{equation}\label{6.31}
    zG_{n,t_r}=H_n\widetilde{F}_{r}-\widetilde{H}_rF_{n},
   \end{equation}
    \begin{equation}\label{6.32}
        H_{n,t_r}=2(H_n\widetilde{G}_r-G_n\widetilde{H}_r).
    \end{equation}
 Equations $(\ref{6.30})-(\ref{6.32})$ imply
    \begin{equation}\label{6.33}
        -B_{n,t_r}+[\widetilde{B}_r,B_n]=0.
    \end{equation}
 \end{lem6.3}
\textbf{Proof.}~~Differentiating both sides of (\ref{6.23}) with
respect to $t_r$ leads to
  \begin{equation}\label{6.34}
    (m_+ - m_-)_{t_r}=-2k(P) F_{n,t_r}F_{n}^{-2}.
  \end{equation}
On the other hand, by (\ref{6.22}), (\ref{6.23}) and (\ref{6.25}),
the left-hand side of (\ref{6.34}) equals to
   \begin{eqnarray}\label{6.35}
    m_{+,t_r}-m_{-,t_r}&=&
    2\widetilde{G}_r(m_+ - m_-)-z^{-\frac{1}{2}}\widetilde{F}_{r}
    (m_+^2-m_-^2)
     \nonumber \\
    &=&
    4k(P)(\widetilde{G}_rF_{n}-\widetilde{F}_{r}G_n)F_{n}^{-2}.
   \end{eqnarray}
Combining (\ref{6.34}) with (\ref{6.35}) yields (\ref{6.30}).
Similarly, Differentiating both sides of (\ref{6.22}) with respect
to $t_r$ gives
  \begin{equation}\label{6.36}
    (m_+ + m_-)_{t_r}=2z^{\frac{1}{2}}
    (G_{n,t_r}F_{n}-G_nF_{n,t_r})F_{n}^{-2},
  \end{equation}
Meanwhile, by (\ref{6.22}), (\ref{6.23}) and (\ref{6.25}), the
left-hand side of (\ref{6.36})  equals to
  \begin{eqnarray}\label{6.37}
  m_{+,t_r}+m_{-,t_r}&=&
   2\widetilde{G}_r(m_+ + m_-)
   -z^{-\frac{1}{2}}\widetilde{F}_{r}(m_+^2+m_-^2)
   -2z^{-\frac{1}{2}}\widetilde{H}_r
    \nonumber \\
   &=&
   -2z^{\frac{1}{2}}G_nF_{n}^{-2}F_{n,t_r}
   +2z^{-\frac{1}{2}}F_{n}^{-1}(-\widetilde{H}_rF_{n}+\widetilde{F}_{r}H_n).
  \end{eqnarray}
Thus, (\ref{6.31}) clearly  follows by (\ref{6.36}) and
(\ref{6.37}). Next, differentiating both sides of (\ref{6.24}) with
respect to $t_r$ yields
   \begin{equation}\label{6.38}
    (m_+m_-)_{t_r}=(H_{n,t_r}F_n-H_nF_{n,t_r})F_n^{-2}.
   \end{equation}
By using (\ref{6.22}), (\ref{6.24}) and (\ref{6.25}), we compute the
left-hand side of (\ref{6.38}), obtaining
  \begin{eqnarray}\label{6.39}
   (m_+m_-)_{t_r}&=&-z^{-\frac{1}{2}}\widetilde{H}_r(m_+ + m_-)
   +4\widetilde{G}_r m_+ m_-
     \nonumber \\
   &&
   -z^{-\frac{1}{2}}\widetilde{F}_r m_+ m_-(m_+ + m_-)
    \nonumber \\
   &=&
   2(\widetilde{G}_rH_n-\widetilde{H}_rG_n)F_n^{-1}
   -H_nF_{n,t_r}F_n^{-2},
  \end{eqnarray}
and hence (\ref{6.32}) holds. Finally, a direct calculation shows
that (\ref{6.30})-(\ref{6.32})
are equivalent to (\ref{6.33}). \quad $\square$ \\

The properties of the $x$-motion and $t_r$-motion of the poles
$P_j(x,t_r)$ now reads as follows.

\newtheorem{the6.4}[lem6.1]{Theorem}
 \begin{the6.4}
  Assume that $(\ref{6.4})$ and $(\ref{6.5})$ hold. Then, for every
  $j=1,\ldots,n$,
  \begin{equation}\label{6.40}
      P_{j,x}(x,t_r))=\frac{2 (-1)^n k(P_j(x,t_r))
      \sqrt{P_j(x,t_r)} \prod_{i=1}^n
      P_i(x,t_r)}
      {k(0) \prod_{j \neq i} (P_j(x,t_r)-P_i(x,t_r))},
     \end{equation}
  and
   \begin{equation}\label{6.41}
     \begin{split}
      P_{j,t_r}(x,t_r))&=\frac{2 (-1)^n k(P_j(x,t_r))
      \widetilde{F}_r(P_j(x,t_r)) \prod_{i=1}^n
      P_i(x,t_r)}
      {k(0) \sqrt{P_j(x,t_r)} \prod_{j \neq i}
      (P_j(x,t_r)-P_i(x,t_r))} \\
      &
      =\frac{\widetilde{F}_r(P_j(x,t_r))}{P_j(x,t_r)}
        P_{j,x}(x,t_r).
      \end{split}
     \end{equation}
 \end{the6.4}
\textbf{Proof.}~~It suffices to focus on (\ref{6.41}), since the
proof procedure for (\ref{6.40}) is analogous to (\ref{5.7}).
Differentiating on both sides of (\ref{5.12}) with respect to $t_r$
yields
   \begin{equation}\label{6.42}
    F_{n,t_r}(P_j)=-\frac{P_{j,t_r}}{\sqrt{y}}
    \prod_{\scriptstyle i \neq j} (P_j-P_i).
   \end{equation}
On the other hand, considering (\ref{6.18}), we compute (\ref{6.30})
at $P_j$,
   \begin{equation}\label{6.43}
    F_{n,t_r}(P_j)=2\widetilde{F}_r(P_j)G_n(P_j)
    =2\widetilde{F}_r(P_j) \frac{k(P_j)}{\sqrt{P_j}}.
   \end{equation}
Hence, combining (\ref{6.42}) and (\ref{6.43}) leads to
(\ref{6.41}). \quad $\square$ \\

\newtheorem{rem6.5}[lem6.1]{Remark}
 \begin{rem6.5}
A closer look at Theorem $6.4$ reveals that the pole motion (both
for the $x$-motion and the $t_r$-motion) can be determined by
solving only first order differential equations
 \begin{equation*}
\frac{\partial P_j}{\partial x}=U_1(P_j) \quad \textrm{and} \quad
\frac{\partial P_j}{\partial t_r}=U_2(P_j),
 \end{equation*}
 where $U_1$ and $U_2$ are bounded continuous functions defined on $\mathbb{R}$.
 \end{rem6.5}

Now we shall provide the algebro-geometric formula for
time-dependent HS solutions $u(x,t_r)$.

\newtheorem{the6.6}[lem6.1]{Theorem}
 \begin{the6.6}
 Assume that $(\ref{6.4})$ and $(\ref{6.5})$ hold. Then the $r$-th
 order HS equation $(\ref{4.11})$ admits a global solution $u(x,t_r)$ of
 algebro-geometric type, when the initial condition
 $u(x,t_0)=u_0(x)$ is given by the solution of the stationary HS
 equation of order $n$. The function
 $y_0(x)=y(x,t_0)=u_{xx}(x,t_0)/2$ lies in the isospectral class
 given by Hypotheses $2.4$ for the Sturm-Liouville operator $L \psi_1=
 \psi_{1,xx}=-zy\psi_1$.

 In particular,
   \begin{equation}\label{6.44}
    u(x,t_r)=\sum_{j=1}^n \frac{1}{P_j(x,t_r)}
            -\sum_{m=1}^{2n} \frac{1}{z_m},
   \end{equation}
 where the pole motion is completely determined from $(\ref{6.40})$
 and $(\ref{6.41})$.

 Moreover, for every $t_r \in \mathbb{R}$, the function
 $y_{t_r}(x)=y(x,t_r)=u_{xx}(x,t_r)/2$ lies in the same isospectral
 class as $y_0(x)$. That is, time evolution of the solutions of the
 $r$-th order HS equation define densities lying in the same
 isospectral class, and this isospectral class depends only on the
 initial data.
 \end{the6.6}
\textbf{Proof.}~~The proof of (\ref{6.44}) is analogous to Theorem
5.5. To show that all the solutions define densities lying in the
same isospectral class, it is sufficient to observe that the
$t_r$-evolution of the poles $P_j(x,t_r)$ implies that, if we start
from a time $\tilde{t}_r$, then the motion $t_r \mapsto P_j(x,t_r)$
remains in the resolvent interval $[z_{2j-1},z_{2j}]~
(j=1,\ldots,n)$. Hence, we can apply Theorem 2.6 to conclude the
result. \quad $\square$

\section{Jacobian flows and pole motion on the generalized Jacobian}
In this section we move our attention to the generalized Jacobian
variety $J_0(\mathcal{R})$ of the Riemann surface $\mathcal{R}$. The
aim is to make clearer the structure of the solutions $u(x,t_r)$
given in (\ref{6.44}).

We first give a short description of what a generalized Jacobian is,
for more details, one can refer to \cite{17,34}.

Let $\mathcal{R}$ denote a hyperelliptic Riemann surface of genus
$n$ with a standard homology basis $a_i,b_i$ $(i=1,\ldots,n)$, and
$dw_1,\ldots,dw_n$ be a basis of normalized holomorphic
differentials on $\mathcal{R}$. The normalized differentials $dw_i$,
means
   \begin{equation*}
    \int_{a_i} dw_j=\delta_{i,j}.
   \end{equation*}
We construct a singular Riemann surface $\mathcal{R}_0$, by pinching
a nonzero homology cycle at a ramification point $Q_0$ on
$\mathcal{R}$.
Roughly speaking, a standard basis of holomorphic
differentials is no longer sufficient to describe the structure of
the Jacobian variety connecting to this new surface $\mathcal{R}_0$:
such a Jacobian variety is called generalized Jacobian variety, and
we denote it by $J_0(\mathcal{R})$.

To solve inverse problems on generalized Jacobian variety, we need
the normalized differential of the second kind $dw_{Q_0}^{(2)}$ on
$\mathcal{R}$ having a double pole at $Q_0$, with principal part
$\lambda_{Q_0}^{-2} d\lambda_{Q_0}$, The normalization means
  \begin{equation*}
   \int_{a_i} dw_{Q_0}^{(2)} = 0,
   \qquad (i=1,\ldots,n).
  \end{equation*}

Let us give a more precise definition of $J_0(\mathcal{R})$. We take
the set $Symm^{n+1}(\mathcal{R})$ of unordered $n+1$-tuples of
points on $\mathcal{R}$, such tuples are called divisors. Two
divisors $\mathcal{D}_1$ and $\mathcal{D}_2$ are identified if
$\mathcal{D}_1-\mathcal{D}_2$ is the divisor of a meromorphic
function on $\mathcal{R}_0$. For a more detailed treatment we refer
to \cite{38}. The map $\tilde{I}_0$
   \begin{equation*}
    \Big(
    \sum_{i=0}^n \int_{P^\ast}^{P_i}
    (dw_{Q_0}^{(2)}, dw_1,\ldots,dw_n )
    \Big)
    \in \mathbb{C}^{n+1}
   \end{equation*}
sends each $(P_0,P_1,\ldots,P_n) \in Symm^{n+1}(\mathcal{R})$ to a
complex vector $(z_0,s) \in \mathbb{C} \times \mathbb{C}^n$, where
$P^\ast$ is a fixed initial point different from the ramification
points of $\mathcal{R}$.

Let $\Lambda_0 \subset \mathbb{C}^{n+1}$ be the $\mathbb{Z}$-lattice
spanned by all vectors of the form
  \begin{equation*}
   \int_\alpha(dw_{Q_0}^{(2)},dw_1,\ldots,dw_n),
   \quad \alpha=a_0,a_1,\ldots,a_n,b_1,\ldots,b_n.
  \end{equation*}
Clearly $\int_{a_0} dw_{Q_0}^{(2)}=0$ and $\int_{a_0} dw_i =0$
$(i=1,\ldots,n)$, where $a_0$ is a sufficiently small simple closed
curve centered at $Q_0 \in \mathcal{R}$ \footnote{$a_0$ is required
to bound a disc in $\mathcal{R}$ centered at $Q_0$, and the support
of $a_0$ is required to be disjoint from the supports of $a_i$ and
$b_i$ $(i=1,\ldots,n)$.}. It turns out that the rank of $\Lambda_0$
is $2n$.
The generalized Jacobian $J_0(\mathcal{R})$ is defined by,
$J_0(\mathcal{R})=\mathbb{C}^{n+1}/\Lambda_0$. It can be shown that
$J_0(\mathcal{R}) \equiv \mathbb{C}^\ast \times \mathbb{C}^n $. For
more details about generalized Jacobian, one can see \cite{17,38}.

Given this construction, we can solve the Jacoby inversion problem
when in presence of nonholomorphic differentials. We define the
generalized Riemann Theta function
   \begin{equation*}
    \Theta_0(\underline{z})=z_0 \Theta(s) +\partial_{\omega_n} \Theta(s),
    \qquad
    \underline{z}=(z_0,s)\in \mathbb{C} \times \mathbb{C}^n,
   \end{equation*}
where $\Theta(s)$ is the classical Riemann Theta function associated
to the Riemann surface $\mathcal{R}$ of genus $n$, and $\omega_n$ is
defined in (\ref{8.5}). From the standard theory, we know that every
symmetric function of divisors $(P_0,P_1,\ldots,P_n)$ can be
expressed in terms of a Theta quotient.

We define the generalized Abel map, $I_0: Symm^{n+1}(\mathcal{R})
\rightarrow J_0(\mathcal{R})=\mathbb{C}^{n+1} / \Lambda_0$, by
    \begin{equation*}
        I_0(P_0,P_1,\ldots,P_n)=
        \Big(
        \sum_{i=0}^n \int_{P^\ast}^{P_i}
        (dw_{Q_0}^{(2)},dw_1,\ldots,dw_n)
        \Big)
        \quad \mathrm{mod} ~ \Lambda_0.
    \end{equation*}
Then it is clear that by setting $P_0=P^\ast$, the generalized Abel
map $I_0$ defines an isomorphism between the space
$Symm^n(\mathcal{R})$ and a non compact subvariety of
$J_0(\mathcal{R})$, given by the locus of the zeros of the
generalized Theta function $\Theta_0$. That is, $I_0$ sends divisors
of degree $n$ into vectors in $J_0(\mathcal{R})$ of the form
     \begin{equation*}
        I_0(P_1,\ldots,P_n)=\tilde{z}=
        \Big(
        \sum_{i=1}^n \int_{P^\ast}^{P_i}
        (dw_{Q_0}^{(2)},dw_1,\ldots,dw_n)
        \Big)
        \in \mathbb{C}^{n+1}
    \end{equation*}
such that $\Theta_0(\tilde{z}-\Delta_0)=0$, where $\Delta_0$ is the
generalized vector of Riemann constants.

The subvariety $I_0(Symm^n(\mathcal{R})) \subset J_0(\mathcal{R})$
is denoted by $\Upsilon_0$. It turns out that every symmetric
rational function of divisors of degree $n$ can be expressed as the
restriction of the corresponding function of divisors of degree
$n+1$ by setting $P_0=P^\ast$, hence it can be written as a theta
quotient. We will study these facts in more detail in Section 8.

Now we turn to our case, $\mathcal{R}$ is the Riemann surface
described by the following algebraic relation,
    \begin{equation*}
        w^2(z)-z\prod_{m=1}^{2n}(z-z_m)=0,
    \end{equation*}
$Q_0=0$, and the corresponding nonholomorphic differential is
$dw_0=dw_{Q_0}^{(2)}$. Let $c_j=\pi^{-1}[z_{2j-1},z_{2j}]$
$(j=1,\ldots,n)$, each $c_j$ is a simple closed curve in
$\mathcal{R}$. The product $c_1 \times c_2 \times \ldots \times c_n$
is a real $n$-torus, and embeds into $Symm^n(\mathcal{R})$, and
hence into $\Upsilon_0$ through the restricted Abel map. Next, we
choose $P_1 \in c_1, \ldots, P_n \in c_n$, then the correspondence
$(P_1,\ldots, P_n) \mapsto \displaystyle \sum_{i=1}^n \int
_{P^\ast}^{P_i}(dw_0,\ldots,dw_{n-1}) $ maps $c_1 \times \ldots
\times c_n$ into a curvilinear parallelogram $\ell \subset
\mathbb{C}^n$, hence the remaining coordinate $\displaystyle
\sum_{i=1}^n \int_{P^\ast}^{P_i} dw_n$ can be viewed as a nonlinear
transcendental function on $\ell$ of $\displaystyle \sum_{i=1}^n
\int_{P^\ast}^{P_i} (dw_0,dw_1,\ldots,dw_{n-1}),$ such that
$\displaystyle \sum_{i=1}^n \int_{P^\ast}^{P_i}
(dw_0,dw_1,\ldots,dw_{n-1},dw_n) \in \Upsilon_0$.

Next, we will study the nature of the motion of the poles
$P_j(x,t_r)$ on $J_0(\mathcal{R})$. To this aim, let
$\mathcal{P}=(P_1,\ldots,P_n) \in \mathbb{C}^n$ be an arbitrary
$n$-tuple of distinct points. We introduce the following symmetric
functions:
   \begin{equation}\label{7.1}
   \varsigma_i(\mathcal{P})=(-1)^i \sum_{\ell \in \Lambda_i}
   P_{\ell_1} \ldots P_{\ell_i},
   \qquad \ell=(\ell_1,\ldots,\ell_i),
   \quad  1 \leq i \leq n,
   \end{equation}
where $ \Lambda_i=\{\ell \in \mathbb{N}^i ~ | ~ 1 \leq \ell_1 <
\ldots < \ell_i \leq n\}$;
   \begin{equation}\label{7.2}
   \sigma_i^{(j)}(\mathcal{P})=(-1)^i \sum_{\ell \in \Lambda_i^{(j)}}
   P_{\ell_1} \ldots P_{\ell_i},
   \qquad \ell=(\ell_1,\ldots,\ell_i),
   \quad  1 \leq i \leq n-1,
   \end{equation}
where $\Lambda_i^{(j)}=\{\ell \in \mathbb{N}^i ~ | ~ 1 \leq \ell_1 <
\ldots < \ell_i \leq n,~\ell_k \neq j\}$.\\
The general form of Lagrange's interpolation theorem then reads as
follows.
\newtheorem{the7.1}{Theorem}[section]
 \begin{the7.1} [Lagrange interpolation formula]
  Let $\mathcal{P}=(P_1,\ldots,P_n) \in \mathbb{C}^n$ be a $n$-tuple of
  distinct points. Then for every $k=1,\ldots,n+1,$ $i=0,\ldots,n-1$
  and $j=1,\ldots,n$, we have
   \begin{equation}\label{7.3}
    \sum_{j=1}^n \frac{P_j^{k-1}}{\prod_{s \neq j} (P_j-P_s)}
    \sigma_i^{(j)}(\mathcal{P})=\delta_{k,n-i}
    -\varsigma_{i+1}(\mathcal{P})    \delta_{k,n+1}.
   \end{equation}
 \end{the7.1}

The simplest Lagrange interpolation formula reads in the case $i=0$,
  \begin{equation*}
    \sum_{j=1}^n \frac{P_j^{k-1}}{\prod_{s \neq j} (P_j-P_s)}
    =\delta_{k,n}, \quad k=1,\ldots,n.
  \end{equation*}

For use in present paper, we recall some important properties of the
functions $\sigma_i^{(j)}$ and $\varsigma_i$. These results can be
found in \cite{12}, here we omit the proofs.

\newtheorem{lem7.2}[the7.1]{Lemma}
  \begin{lem7.2}
  Let $\mathcal{P}=(P_1,\ldots,P_n) \in \mathbb{C}^n$ be a $n$-tuple of
  distinct points. Then
    \begin{eqnarray*}
     &&(i) \quad \varsigma_{i+1}(\mathcal{P})+P_j \sigma_i^{(j)}(\mathcal{P})
        = \sigma_{i+1}^{(j)}(\mathcal{P}),
        \quad i=0,\ldots,n-1,~j=1,\ldots,n.
         \\
   &&(ii) \quad
   \sum_{i=0}^r \varsigma_{r-i}(\mathcal{P}) P_j^i
         =\sigma_r^{(j)}(\mathcal{P}),
         \quad r=0,\ldots,n,~j=1,\ldots,n.
         \\
   &&(i i i) \quad
      \sum_{i=0}^{r-1} \sigma_{r-1-i}^{(j)}(\mathcal{P})z^i
      =\frac{1}{z-P_j}\Big(
      \sum_{i=0}^r \varsigma_{r-i}(\mathcal{P})z^i
      -\sigma_r^{(j)}(\mathcal{P})
      \Big).
    \end{eqnarray*}
  \end{lem7.2}

We define the (non normalized) differentials
      \begin{equation*}
       d\omega_{s-1}=\frac{z^{s-2}}{ 2 \sqrt{z}~ k(z) }dz,
       \quad s=1,\ldots,n.
      \end{equation*}
Note that $d\omega_0$ is a differential of the second kind, having a
double pole at $0$, with coefficient $(1/\sqrt{\prod z_m})$ of
corresponding principal part.

Moreover, let
   \begin{equation*}
    \omega_{s-1}(x,t_r)=\sum_{i=1}^n \int_{P^\ast}^{P_i(x,t_r)}
    d\omega_{s-1}, \quad s=1,\ldots,n.
   \end{equation*}
Then differentiating each $\omega_{s-1}$ with respect to $x$, we
have
   \begin{equation}\label{7.4}
    \frac{\partial \omega_{s-1}(x,t_r)}{\partial x}=
    \sum_{j=1}^n \frac{P_j^{s-2}(x,t_r)}{2 \sqrt{P_j(x,t_r)} k(P_j(x,t_r))}
    \frac{\partial P_j(x,t_r)}{\partial x}.
   \end{equation}
Substituting (\ref{6.40}) into (\ref{7.4}) and taking account into
Theorem 7.1 yields
   \begin{equation}\label{7.5}
    \frac{\partial \omega_{s-1}(x,t_r)}{\partial x}=
    \frac{-1}{k(0)}
    \sum_{j=1}^n\frac{P_j^{s-1}(x,t_r) \sigma_{n-1}^{(j)}(\mathcal{P})}
    {\prod_{i \neq j}(P_j(x,t_r)-P_i(x,t_r))}
    =\frac{-1}{k(0)}\delta_{s,1},
    \end{equation}
which implies the following relation:
   \begin{equation}\label{7.6}
   \frac{\partial \omega_{s-1}(x,t_r)}{\partial x}=
     \begin{cases}
      \displaystyle -\frac{1}{k(0)},& s=1, \\
      0, & s=2,\ldots,n.
     \end{cases}
   \end{equation}
Hence we obtain
    \begin{equation}\label{7.7}
     \omega_{s-1}(x,t_r)=
      \begin{cases}
       c_0(t_r)-\displaystyle \frac{1}{k(0)}x, & s=1, \\
       c_{s-1}(t_r), & s=2,\ldots,n.
      \end{cases}
    \end{equation}
This result shows that the $x$-motion is constant with respect to
$n-1$ coordinates, while it is linear with respect to $x$ on the
remaining coordinate, which corresponds to the differential of the
second kind.

Now we investigate the $t_r$-motion of the poles $P_j(x,t_r)$.
Recall (\ref{5.12}):
   \begin{equation*}
    F_n(z)=\frac{1}{\sqrt{y(x,t_r)}}
    \prod_{i=1}^n (z- \pi (P_i(x,t_r))).
   \end{equation*}
By the construction of $\widetilde{F}_r(z)$, we know that for every
$r=1,\ldots,n$, the polynomial $\widetilde{F}_r(z)$ can be obtained
by truncating the polynomial $F_n(z)$ at the degree $r$, that is
   \begin{equation}\label{7.8}
    \widetilde{F}_r(z)=F_n(z)-z^{r+1}
    \left[\frac{F_n(z)}{z^{r+1}} \right]_p,
   \end{equation}
where $[~~]_p$ denotes the polynomial part. Hence, we conclude that
  \begin{equation}\label{7.9}
   \widetilde{F}_r(P_j(x,t_r))=-P_j^{r+1}(x,t_r)
    \left[\frac{F_n(P_j(x,t_r))}{z^{r+1}} \right]_p
    =-\frac{P_j^{r+1}(x,t_r)}{\sqrt{y(x,t_r)}}
    \sigma_{n-r-1}^{(j)}(\mathcal{P}).
  \end{equation}
Inserting (\ref{7.9}) into (\ref{6.41}) yields
  \begin{eqnarray}\label{7.10}
  P_{j,t_r}(x,t_r)&=&-\frac{P_j^{r}(x,t_r)}{\sqrt{y(x,t_r)}}
    \sigma_{n-r-1}^{(j)}(\mathcal{P}) P_{j,x}(x,t_r)
     \nonumber \\
    &=&
    2\frac{P_j^r(x,t_r) \sigma_{n-r-1}^{(j)}(\mathcal{P})
    k(P_j(x,t_r)) \sqrt{P_j(x,t_r)}}
    {\displaystyle \prod_{i \neq j} (P_j(x,t_r)-P_i(x,t_r))}.
  \end{eqnarray}
As before, differentiating each $\omega_{s-1}$ with respect to $t_r$
gives
    \begin{equation}\label{7.11}
    \frac{\partial \omega_{s-1}(x,t_r)}{\partial t_r}=
    \sum_{j=1}^n \frac{P_j^{s-2}(x,t_r)}{2 \sqrt{P_j(x,t_r)} k(P_j(x,t_r))}
    \frac{\partial P_j(x,t_r)}{\partial t_r},
    \end{equation}
and insertion of (\ref{7.10}) into (\ref{7.11}) yields the beautiful
relation
     \begin{equation}\label{7.12}
       \frac{\partial \omega_{s-1}(x,t_r)}{\partial t_r}=
       \sum_{j=1}^n \frac{P_j^{s+r-2}(x,t_r) \sigma_{n-r-1}^{(j)}(\mathcal{P})}
       {\displaystyle \prod_{i \neq j} (P_j(x,t_r)-P_i(x,t_r))},
     \end{equation}
where $s=1,\ldots,n$ and $r=1,\ldots,n-1$.

After a short computation, we arrive at
   \begin{equation}\label{7.13}
   d\omega_{s-1}=\delta_{s-1,1}dt_r
   -\sum_{l=0}^{r-1} \Bigg(
   \sum_{j=1}^n
    \frac{P_j^{s+l-2}\varsigma_{n-l-1}(\mathcal{P})}
    {\displaystyle \prod_{i \neq j} (P_j-P_i)} \Bigg) dt_r.
   \end{equation}

Next, we introduce the notation $\alpha_r=n-r+1$. Then (\ref{7.13})
gives:\\
For every $s=0,\ldots,\alpha_r-1$,
    \begin{equation*}
    d\omega_s=\delta_{s,1} dt_r.
    \end{equation*}
For $k=0,\ldots,r-2$,
    \begin{equation*}
        d\omega_{\alpha_r+k}=-\sum_{h=n-r}^{n-1}
        \Bigg( \sum_{j=1}^n
        \frac{P_j^{k+h} \varsigma_{2n-r-h-1}(\mathcal{P})}
        {\prod_{i \neq j}(P_j-P_i)}\Bigg) dt_r.
    \end{equation*}
If we set
$$dt_{r,h}=\varsigma_{2n-r-h-1}(\mathcal{P}) dt_r$$
and
$$\mathcal{H}_{h+k}(\mathcal{P})=\sum_{j=1}^n
\frac{P_j^{k+h}}{\prod_{i \neq j}(P_j-P_i)},$$
then we obtain
   \begin{equation*}
   d\omega_{\alpha_r+k}=-\sum_{h=n-r}^{n-1}
   \mathcal{H}_{h+k}(\mathcal{P})dt_{r,h}.
   \end{equation*}
As long as $k+h$ does not reach the value $n-1$, then
$\mathcal{H}_{h+k}(\mathcal{P})=0$. Hence, for every fixed $r$, we
have the following triangular structure for the pole motion:
   \begin{eqnarray}\label{7.14}
    d\omega_s(x,t_r)=
    \begin{cases}
     \displaystyle -\frac{1}{k(0)}dx, & s=0, \\
     \\
     dt_r,  &  s=1,\\
     \\
     0,   & s=2,\ldots, \alpha_r-1, \\
     \\
     \displaystyle -\sum_{h=n-r}^{n-1}
   \mathcal{H}_{h+k}(\mathcal{P})dt_{r,h},
   & s=\alpha_r+k, \quad k=0,\ldots, r-2.
    \end{cases}
   \end{eqnarray}
A closer look at (\ref{7.14}) implies the main differences between
our formulas for the pole motion in the generalized Jacobian and
those in \cite{7}:
\begin{enumerate}
\item[1.] In our context, time motion is confined to the holomorphic
coordinates, i.e. $\omega_1(x,t_r)=\omega_1(t_r),\ldots,
\omega_{n-1}(x,t_r)=\omega_{n-1}(t_r)$, while the $x$-motion evolves
only in the meromorphic one, namely $\omega_0(x,t_r)=\omega_0(x)$;
this shows a complete separation between spatial and time motions.

\item[2.] For every $r=1,\ldots,n-1$, the motion is linear with respect
to $x$, with no need of any linearizing change of variables.

\item[3.] The classical one dimensional HS equation corresponds to
the case $r=1$, then the motion on the generalized Jacobian is
remarkably simple:
      \begin{equation*}
      \omega_s(x,t_1)=
      \begin{cases}
      \displaystyle -\frac{1}{k(0)}x+\chi_0, & s=0, \\
       t_1+\chi_1, & s=1,\\
       0, & \textrm{otherwise},
      \end{cases}
      \end{equation*}
where $\chi_0$ and $\chi_1$ are constant phases.
\end{enumerate}

\section{Theta function representation for the solution $u(x,t_r)$
         on the generalized Jacobian}

In our final section we present expressions for all the elementary
symmetric functions of $n$ distinct points on $\mathcal{R}$ in terms
of Riemann theta function. In particular, we obtain the theta
function representation for the solution $u(x,t_r)$ of the HS
hierarchy.

Let
    \begin{equation}\label{8.1}
     d \omega_{k-1}=\frac{z^{k-2}}{2 \sqrt{z}~ k(z)}dz,
     \quad k=1,\ldots,n.
    \end{equation}
We note that for $k >1$, $d \omega_{k-1}$ is a non normalized
holomorphic differential on $\mathcal{R}$. While $k=1$, $d\omega_0$
is a differential of the second kind, having a double pole at $0$
with principal part $(1/\sqrt{\prod z_m})\lambda_0^{-2}d\lambda_0$,
in terms of the local coordinate $\lambda_0=z^{1/2}$ near $0$.

Moreover, we denote by $dw_1,\ldots,dw_g$ a normalized basis of
holomorphic differentials on $\mathcal{R}$, and by $dw_0$ the
normalized differential of the second kind having a double pole at
$0$ with principal part $\lambda_0^{-2}d\lambda_0$, in terms of the
local coordinate $\lambda_0=z^{1/2}$ near $0$.

Let $\mathcal{P}=(P_1,\ldots,P_n)$, where $P_i ~(i=1,\ldots,n)$ are
distinct points on $\mathcal{R}$. For our convenience, we write
$P_i$ instead of $\pi(P_i)$ if no confusion arises. Define the
variables $\alpha_1,\ldots,\alpha_n$, such that
    \begin{equation}\label{8.2}
     \frac{\partial P_j}{\partial \alpha_k}=
     \frac{2 P_j^{3/2} \sigma_{n-k}^{(j)}(\mathcal{P}) k(P_j)}
     {\prod_{i\neq j}(P_j-P_i)},
     \qquad j,k=1,\ldots,n,
    \end{equation}
where $P_j=P_j\,(\alpha_1,\ldots,\alpha_n)$. Next, we give the
explicit formulas for $\alpha_1,\ldots,\alpha_n$. Combining
(\ref{8.1}) and (\ref{8.2}), we obtain
    \begin{eqnarray}\label{8.3}
    \frac{\partial }{\partial \alpha_k}(\omega_{k-1})&=&
     \frac{\partial }{\partial \alpha_k}
     \left(
     \sum_{j=1}^n \int_{P^\star}^{P_j\,(\alpha_1,\ldots,\alpha_n)}
      \frac{z^{k-2}}{2 \sqrt{z}~ k(z)}dz
     \right) \nonumber \\
     &=&
     \sum_{j=1}^n \frac{P_j^{k-2}}{2 \sqrt{P_j}~ k(P_j)}
     \frac{\partial P_j}{\partial \alpha_k}
     =\sum_{j=1}^n\frac{P_j^{k-1} \sigma_{n-k}^{(j)}(\mathcal{P})}
     {\prod_{i\neq j}(P_j-P_i)}
     \nonumber \\
    &=&
    \delta_{k,k}-\varsigma_{n-k+1}\delta_{k,n+1}=1.
    \end{eqnarray}
Hence, we have
    \begin{equation}\label{8.4}
    \alpha_k-\alpha_k^0=\sum_{j=1}^n
    \int_{P^\ast}^{P_j}d\omega_{k-1}=\omega_{k-1},
    \quad k=1,\ldots,n.
    \end{equation}
Next, we introduce the additional function
    \begin{equation}\label{8.5}
    \alpha_{n+1}=\sum_{j=1}^n
    \int_{P^\ast}^{P_j(\alpha_1,\ldots,\alpha_n)}
    \frac{z^{n-1}}{2\sqrt{z}~k(z)}dz
    =\omega_n(\mathcal{P}).
    \end{equation}
Differentiating (\ref{8.5}) with respect to $\alpha_k$ on both sides
gives
   \begin{equation}\label{8.6}
   \frac{\partial \alpha_{n+1}}{\partial \alpha_k}=
   \sum_{j=1}^n\frac{P_j^n}{\prod_{i\neq j}(P_j-P_i)}
   \sigma_{n-k}^{(j)}(\mathcal{P})
   =\delta_{n+1,k}-\varsigma_{n-k+1}(\mathcal{P})
   =-\varsigma_{n-k+1}(\mathcal{P}).
   \end{equation}
Equation (\ref{8.6}) shows that all the symmetric functions of the
poles $P_1,\ldots,P_n$, which are restrictions of symmetric
functions of $n+1$ points in $J_0(\mathcal{R})$, can be determined
by differentiating the $n$-th non normalized Abel coordinate with
respect to the $(k-1)$-th one.

There is a normalizing matrix $D\in Sl(n,\mathbb{C})$ such that
  \begin{equation}\label{8.7}
  \underline{\omega}=D\underline{w},
  \end{equation}
where $\underline{\omega}$ and $\underline{w}$ denote the column
vectors of the Abel coordinates $(\omega_1,\ldots,\omega_n)$ and
$(w_1,\ldots,w_n)$ respectively. Moreover, there are constants
$\eta_1,\ldots,\eta_n \in \mathbb{C}$, such that
   \begin{equation}\label{8.8}
   dw_0=k(0)d\omega_0+\sum_{i=1}^n\eta_idw_i,
   \end{equation}
where
   \begin{equation*}
   \eta_i=-k(0)\int_{a_i}d\omega_0.
   \end{equation*}
For $s,k=1,\ldots,n$, a direct calculation yields
   \begin{equation}\label{8.9}
   \frac{\partial \alpha_s}{\partial \alpha_k}=
   \sum_{j=1}^n\frac{P_j^{s-1} \sigma_{n-k}^{(j)}(\mathcal{P})}
   {\prod_{i\neq j}(P_j-P_i)}=\delta_{s,k}.
   \end{equation}
We denote by $D=(\gamma_{rs})$ and $D^{-1}=(\beta_{rs})$. Then from
(\ref{8.9}) we infer
   \begin{equation}\label{8.10}
   \frac{\partial w_s}{\partial \alpha_k}=\sum_{j=1}^n \beta_{sj}
   \frac{\partial \omega_j}{\partial \alpha_k}
   =\sum_{j=1}^{n-1}\beta_{sj}
   \frac{\partial \alpha_{j+1}}{\partial \alpha_k}
   +\beta_{sn}\frac{\partial \alpha_{n+1}}{\partial \alpha_k}
   =\beta_{s,k-1}-\beta_{sn}\varsigma_{n-k+1}(\mathcal{P}),
   \end{equation}
where $\beta_{s0}=0$.

Now we are intend to determine the dependence of $ \displaystyle
\frac{\partial \omega_n}{\partial \alpha_k}$ with respect to the
classical Riemann theta function  $\Theta(s)$, where
$s=s(\mathcal{P})=I(P_1,\ldots,P_n)$ and $I:Symm^n(\mathcal{R})
\rightarrow J(\mathcal{R})$ denotes the
standard Abel map.\\
Since $ w_0(P_1,\ldots,P_n)=-\partial_{\omega_n} \mathrm{ln} \,
\Theta(s(\mathcal{P}))$ (see \cite{7,9}), then combining (\ref{8.7})
and (\ref{8.8}) yields
   \begin{equation}\label{8.11}
   \omega_n=\sum_{s=1}^n \gamma_{ns}w_s=\sum_{s=1}^{n-1}\gamma_{ns}w_s
   +\frac{\gamma_{nn}}{\eta_n} \left(
   -\partial_{\omega_n} \mathrm{ln} \, \Theta(s(\mathcal{P}))
   -k(0)\alpha_1-\sum_{i=1}^{n-1}\eta_iw_i
   \right).
   \end{equation}
Then differentiating (\ref{8.11}) with respect to $\alpha_k$, after
some computations, we arrive at
  \begin{equation}\label{8.12}
  \left(\sum_{s=1}^n \beta_{sn}\eta_s \right)
  \varsigma_{n-k+1}(\mathcal{P})
  =\left(\sum_{s=1}^n \beta_{s,k-1}\eta_s \right)
  +k(0)\delta_{1,k}+ \frac{\partial^2}{\partial \alpha_k \omega_n}
  \mathrm{ln}\Theta(s(\mathcal{P})).
  \end{equation}

Introducing the notation
\begin{equation*}
\xi_k=\left(\sum_{s=1}^n\beta_{sk}\eta_s \right), \quad
k=1,\ldots,n.
\end{equation*}
Next, we are intend to make the clear the meaning of the constants
$\xi_k$. For this purpose, consider the $(n+1) \times (n+1)$ matrix
   \begin{equation*}
   D_0=
   \left(
     \begin{array}{cccc}
       1/k(0) & -\eta_1/k(0) & \ldots & -\eta_n/k(0) \\
       0 &  &  &  \\
       \vdots &  & D &  \\
       0 &  &  &  \\
     \end{array}
   \right).
   \end{equation*}
It is clear that $D_0$ is a normalizing matrix in the sense that
  \begin{equation*}
  \left(
    \begin{array}{c}
      \omega_0 \\
      \omega_1 \\
      \vdots \\
      \omega_n \\
    \end{array}
  \right)
  =
  D_0
  \left(
    \begin{array}{c}
      w_0 \\
      w_1 \\
      \vdots \\
      w_n \\
    \end{array}
  \right).
  \end{equation*}
By a short computation, one can infer that the first row of the
inverse matrix $D_0^{-1}$ is the vector $(k(0),\xi_1,\ldots,\xi_n)$.
This implies that the constants $\xi_k$ are those complex numbers
such that
   \begin{equation}\label{8.13}
    dw_0=k(0)d\omega_0+\sum_{k=1}^n\xi_kd\omega_k.
   \end{equation}

Based on the above analysis, we have the following result.

\newtheorem{lem8.1}{Lemma}[section]
 \begin{lem8.1}
  Let $\mathcal{P}=(P_1,\ldots,P_n)$ be a $n$-tuple of distinct
  points of $\mathcal{R}$. Then for every $k=1,\ldots,n$,
    \begin{equation}\label{8.14}
    \varsigma_{n-k+1}(\mathcal{P})=\frac{1}{\xi_n}
    \left( \xi_{k-1}+k(0)\delta_{1,k}+
    \frac{\partial^2 }{\partial \omega_{k-1} \omega_n}
    \mathrm{ln}\Theta(s(\mathcal{P}))
    \right),
    \end{equation}
 where $s(\mathcal{P})=I(P_1,\ldots,P_n)$,
 $I:Symm^n(\mathcal{R}) \rightarrow J(\mathcal{R})$ is
 the standard Abel map, and the constants $\xi_k$
 satisfy $(\ref{8.13})$, with $\xi_0=0$, $\alpha_k=\omega_{k-1}$.
 \end{lem8.1}

For every $n$-tuple $\mathcal{P}=(P_1,\ldots,P_n)$ of distinct
points of $\mathcal{R}$, the symmetric rational function $\varrho:
Symm^n(\mathcal{R}) \rightarrow \mathbb{C}$ $ \displaystyle
:\mathcal{P} \mapsto \prod_{i=1}^n \pi(P_i)$ is the restriction to
$Symm^n(\mathcal{R})$, obtained by taking $P_0=P^\ast$, of a
symmetric rational function $\rho: Symm^{n+1}(\mathcal{R})
\rightarrow \mathbb{C}$. Following \cite{37}, every symmetric
rational function $\rho$ on $Symm^{n+1}(\mathcal{R})$ defines a
meromorphic function on $J_0(\mathcal{R})$. Hence, from
(\ref{8.14}), after some manipulations, we obtain the formula for
the symmetric function
    \begin{equation}\label{8.15}
    \varsigma_n(\mathcal{P})=\prod_{i=1}^n P_i
    =\gamma \frac{\displaystyle \Theta_0^2 \left(I_0(P_1,\ldots,P_n)-
    \int_{P^\ast}^{0} (w_0,w_1,\ldots,w_n)-\Delta_0 \right)}
    {\displaystyle \Theta_0^4 \left(I_0(P_1,\ldots,P_n)-
    \int_{P^\ast}^{\infty} (w_0,w_1,\ldots,w_n)-\Delta_0 \right)},
    \end{equation}
where $\gamma$ is a constant depending only on the choice of the
base point $P^\ast$ and the genus $n$ of $\mathcal{R}$, and
$\Delta_0$ is the generalized vector of Riemann constants. For
notational simplicity, we denote the right-hand side of (\ref{8.15})
by $(-1)^n\hat{\Theta}_0(\mathcal{P})$.

We note that (\ref{6.44}) can be rewritten as
   \begin{equation}\label{8.16}
   u(x,t_r)=-\frac{\varsigma_{n-1}(\mathcal{P}(x,t_r))}
   {\varsigma_{n}(\mathcal{P}(x,t_r))}
   -\sum_{i=1}^{2n}\frac{1}{z_i},
   \end{equation}
where $\mathcal{P}(x,t_r)=(P_1(x,t_r),\ldots,P_n(x,t_r))$.\\

Our main result, the theta function representation of the
algebro-geometric  solution $u(x,t_r)$ for the HS hierarchy now
follows from the material prepared above.

\newtheorem{the8.2}[lem8.1]{Theorem}
 \begin{the8.2}
  The solution $u(x,t_r)=u(P_1(x,t_r),\ldots,P_n(x,t_r))$ of the
  $r$-th order HS equation, can be written as the following form
   \begin{equation}\label{8.17}
     u(x,t_r)=\frac{(-1)^{n+1}}{\xi_n \hat{\Theta}_0(\mathcal{P}(x,t_r))}
     \left(
     \frac{\partial^2 }{\partial \omega_1 \omega_n} \mathrm{ln}
     \Theta(s(\mathcal{P}(x,t_r)))
     +\xi_1
     \right)
     -\sum_{i=1}^{2n}\frac{1}{z_i},
     \end{equation}
  where $\mathcal{P}(x,t_r)=(P_1(x,t_r),\ldots,P_n(x,t_r))$,
  $s(\mathcal{P})=I(P_1(x,t_r),\ldots,P_n(x,t_r))$, $\xi_1$
  and $\xi_n$ satisfy $(\ref{8.13})$, and $I:Symm^n(\mathcal{R})
  \rightarrow J(\mathcal{R})$ denotes the standard Abel map.
 \end{the8.2}
\textbf{Proof.}~~The expression (\ref{8.17}) is an immediate
consequence of Lemma 8.1, (\ref{8.15}) and (\ref{8.16}). \quad
$\square$

\newtheorem{rem8.3}[lem8.1]{Remark}
 \begin{rem8.3}
 Theorem $8.2$ shows that the expression $(\ref{8.17})$ depends on the theta
 quotient $(\ref{8.15})$ and the partial derivative with respect to
 the coordinate $\omega_1$ of the logarithm of
 $w_0(P_1(x,t_r),\ldots,P_n(x,t_r))$. Hence, $u(x,t_r)$ can be
 viewed as the restriction of a function, defined on
 $J_0(\mathcal{R})$, to the subvariety $\Upsilon_0$ given by the locus
 of the zeros of the generalized theta function $\Theta_0(z)$.
 \end{rem8.3}

\section{Conclusions}
In this paper, we obtained global solutions of algebro-geometric
type for all the equations of a new commuting hierarchy containing
the Hunter-Saxton equation. As a main tool we used theta function
expressions for all the symmetric functions of points
$P_1,\ldots,P_n \in \mathcal{R}$. Some of these expressions are
apparently new.

On the other hand, the Hunter-Saxton equation belongs to a larger
family called Dym-type equation in \cite{7,8,9},
    \begin{equation*}
    u_{xxt}+2u_xu_{xx}+uu_{xxx}-2\kappa u_x=0,
    \qquad \textrm{$\kappa=$ constant}.
    \end{equation*}
One of these equations is a member of the Dym hierarchy that has
been studied by, amongst others, Kruskal \cite{39}, Cao \cite{40},
Hunter and Zheng \cite{2} and Alber et al. \cite{7,8}.

We remark that although our focus in this paper is on the case
$\kappa=0$, all the arguments presented here can be adapted, with no
obvious modifications, to study the corresponding equation $\kappa
\neq 0$. As it is observed that by substituting
$y=\frac{1}{2}u_{xx}$ into $y=\frac{1}{2}u_{xx}-\frac{1}{2}\kappa$,
then (\ref{4.11}) represents the Dym hierarchy and (\ref{4.12}) will
become the Dym-type equation. The analysis from Section 5 to Section
8 can extend line by line to the Dym hierarchy. Hence, it is trivial
to investigate the algebro-geometric solutions of Dym hierarchy
again.

\section*{Acknowledgments}
 We are deeply indebted to Professor R. Johnson for
 sharing his latest papers.
 This work was supported by grants from the
 National Science Foundation of China (Project No.10971031; No.11271079)
 and the Shanghai Shuguang Tracking Project (Project No.08GG01).


{\small

\begin{thebibliography}{99}
\bibitem{1}J.K. Hunter, R. Saxton, Dynamics of director fields,
 SIAM J. Appl. Math. 51 (1991) 1498-1521.
\bibitem{2}J.K. Hunter, Y.X. Zheng, On a completely integrable nonlinear
hyperbolic variational equation, Physica D. 79 (1994) 361-386.
\bibitem{3} I.M. Krichever, Integration of nonlinear equations by
the methods of algebraic geometry, Funct.Anal.Appl. 11 (1977) 12-26.
\bibitem{4} B.A. Dubrovin, Completely integrable Hamiltonian systems
associated with matrix operators and Abelian varieties,
Funct.Anal.Appl. 11 (1977) 265-277.
\bibitem{5} S.P. Novikov, S.V. Manakov, L.P. Pitaevskii, V.E.
Zakharov, Theory of Solitons, the Inverse Scattering Methods,
Concultants Bureau, New York, 1984.
\bibitem{6} E.D. Belokolos, A.I. Bobenko, V.Z. Enol'skii, A.R. Its,
and V.B. Matveev, Algebro-Geometric Approach to Nolinear Integrable
Equations, Springer, Berlin, 1994.
\bibitem{7} M.S. Alber and Y.N. Fedorov, Algebraic geometrical
sollutions for certain evolution equations and hamiltonian flows on
nonlinear subvarieties of generalized Jacobians, Inverse Problems.
17 (2001) 1017-1042.
\bibitem{8} M.S. Alber, R. Camassa, Y.N. Fedorov, D.D. Holm and
J.E. Marsden, The complex geometry of weak piecewise smooth
solutions of integrable nonlinear PDE's of shallow water and dym
tye, Commun. Math. Phys. 221 (2001) 197-227.
\bibitem{9} M.S. Alber, Y.N. Fedorov, Wave solutions of evolution
equations and hamiltonian flows on nonlinear subvarieties of
generalized Jacobians, J. Phys. A: Math. Gen. 33 (2000) 8409-8425.
\bibitem{10} R. Camassa, D.D. Holm, An integrable shallow water
equation with peaked solitons, Phys.Rev.Lett. 71 (1993) 1661-1664.
\bibitem{11} R. Camassa, D.D. Holm and J.M. Hyman, A new integrable shallow
water equation, Adv.Appl.Mech. 31 (1994) 1-33.


\bibitem{12} F. Gesztesy, H. Holden,
Soliton Equations and Their Algebro-Geometric Solutions, Volume I:
(1+1)-Dimensional Continuous Models, Cambridge Studies in Advanced
Mathematics, Vol. 79, Cambridge University Press, (2003).
\bibitem{13} R. Johnson, L. Zampogni, On the inverse Sturm-Liouville
problem, Discr. Cont. Dynam. Systems. 18 (2007), 405-428.
\bibitem{14} L. Zampogni, On algebro-geometric solutions of the
Camassa-Holm hierarchy, Adavaced Nonlinear Studies. 7 (2007)
345-380.
\bibitem{15} L. Zampogni, On the inverse Sturm-Liouville problem and
the Camassa-Holm equation, Ph.D. Thesis, Universit\`{a} degi Studi
di Firenze, 2006.
\bibitem{16} Y. Hou, E.G. Fan and P. Zhao, The algebro-geometric
solutions for Hunter-Saxton hierarchy, submitted for publication.
\bibitem{17} J.D. Fay, Theta functions on Riemann surfaces, Lecture
Notes n. 352, Springer Verlag, 1973.

\bibitem{18} R. Beals, D.H. Sattinger and J. Szmigielski,
Multipeakons and classical moment problem, Advances in Math. 154
(2000) 229-257.
\bibitem{19}R. Beals, D.H. Sattinger and J. Szmigielski, Inverse scattering
solutions of the Hunter-Saxton equation, Appl. Anal. 78 (2001)
255-269.
\bibitem{20}E.G. Reyes, The soliton content of the Camassa-Holm and
Hunter-Saxton equations,in: A.G. Nikitin, V.M. Boyko, R.O. Popovych
(Eds.), Proceedings of the Fourth International Conference on
Symmetry in Nonlinear Mathematical Physics, in: Proceedings of the
Institute of Mathematics of the NAS of Ukraine, vol. 43, Kyiv, 2002,
pp. 201-208.
\bibitem{21}E.G. Reyes, Pseudo-potentials, nonlocal symmetries,
and integrability of some shallow water equations, Selecta Math.
(N.S.) 12 (2006) 241-270.
\bibitem{22}B. Khesin, G. Misio{\l}ek, Euler equations on homogeneous spaces
and Virasoro orbits, Adv. Math. 176 (2003) 116-144.
\bibitem{23}J. Lenells, Weak geodesic flow and global solutions of
the Hunter-Saxton equation, Discrete Contin. Dyn. Syst. 18 (2007)
643-656.
\bibitem{24}J. Lenells, The Hunter-Saxton equation describes the
geodesic flow on a sphere, J. Geom. Phys. 57 (2007) 2049-2064.
\bibitem{25}A. Bressan, A. Constantin, Global solutions of the
 Hunter-Saxton equation, SIAM J. Math. Anal. 37 (2005) 996-1026.
\bibitem{26} A. Bressan, H. Holden and X. Raynaud, Lipschitz metric
for the Hunter-Saxton equation, J. Math. Pure. Appl. 94 (2010)
68-92.
\bibitem{27}Z. Yin, On the structure of solutions to the
periodic Hunter-Saxton equation, SIAM J. Math. Anal. 36 (2004)
272-283.
\bibitem{28} G.L. Gui, Y. Liu and M. Zhu, On the wave-breaking phenomena
and global existence for the generalized periodic Camassa-Holm
equation, Int. Math. Res. Notices. 10 (2011) 1-46.
\bibitem{29} O.I. Morozov, Contact equivalence of the generalized
Hunter-Saxton equation and the Euler-Poisson equation. Preprint
math-ph/0406016.
\bibitem{30} S. Sakovich, On a Whitham-type equation, Symmetry,
Integrability. Geom: Methods. Appl. (SIGMA) 5 (2009) 1-7.


\bibitem{31} A.S. Fokas, B. Fuchssteiner, Symplectic structures,
their B$\mathrm{\ddot{a}}$cklund transformation and hereditary
symmetries. Phys. D. 4 (1981) 47-66.
\bibitem{32} P. Rosenau, Nonlinear dispersion and compact structures,
 Phys. Rev. Lett. 73 (1994) 737-1741.
\bibitem{33} R. Johnson, L. Zampogni, Description of the algebro-geometric
Sturm-Liouville coefficients, J. Diff. Equ. 244 (2008) 716-740.

\bibitem{34} D. Mumford, Tata Lectures on Theta. Vol. 1,2,3,
Birkh\"{a}user, 1983.

\bibitem{35} R. Carmona, J. Lacroix, Spectral theory of random
schr\"{o}dinger operators, probability and its applications,
Birkh\"{a}user, 1990.
\bibitem{36} E.A. Coddington, N. Levinson, Theory of ordinary
differential equations, McGraw-Hill, 1955.

\bibitem{37} Y. Fedorov, Classical integrable systems and billiards
related to generalized Jacobians, Acta Appl. Math. 55 (1999)
251-301.

\bibitem{38} Y. Hou, E.G. Fan, and P. Zhao, On the inverse
Sturm-Liouville problem of integrable nonlinear PDE's of Dym type,
in preparation.

\bibitem{39} M.D. Kruskal, Nonlinear wave equations. In: J. Moser
(eds.) Dynamical Systems, Theory and Applications, Lecture Notes in
Physics 38, Springer, New York, 1975.

\bibitem{40} C. Cao, Stationary Harry-Dym's equation and its
relation with geodesics on ellipsoid, Acta Math. Sinica. 6 (1990)
35-41.
\end{thebibliography}

}
\end{document}